\documentclass[aps, prx, english, a4paper, floatfix, twocolumn, longbibliography, groupedaddress, superscriptaddress]{revtex4-2}

\usepackage{amsmath}
\usepackage{amssymb}
\usepackage{color}
\usepackage{tikz}
\usetikzlibrary{quantikz}
\usepackage{graphicx, import}
\usepackage[hidelinks]{hyperref}
\usepackage{glossaries}
\usepackage{booktabs}
\usepackage{multirow}
\usepackage{soul}
\usepackage[normalem]{ulem}
\usepackage{algorithm}
\usepackage{algpseudocode}
\usepackage{placeins}
\usepackage[dvipsnames]{xcolor}
\usepackage{braket}
\usepackage[final]{pdfpages}
\makeatletter
\AtBeginDocument{\let\LS@rot\@undefined}
\makeatother

\usepackage{xr}
\begin{document}
\newacronym{dmrg}{DMRG}{density matrix renormalization group}
\newacronym{skqd}{SKQD}{sample-based Krylov quantum diagonalization}
\newacronym{afm}{AFM}{antiferromagnetic}
\newacronym{ml}{ML}{machine learning}
\newacronym{peps}{PEPS}{projected entangled pair states}
\newacronym{obp}{OBP}{operator backpropagation}
\newacronym{hpc}{HPC}{high-performance computing}
\newacronym{qpu}{QPU}{quantum processing unit}
\newacronym{mps}{MPS}{matrix product state}
\newacronym{ipr}{IPR}{inverse participation ratio}
\newacronym{cv}{CV}{cross-validation}
\glsdisablehyper

\title{Learning ground state observables from quantum computing experiments}
\author{Ben Jaderberg}
\affiliation{
IBM Quantum, IBM Research Europe, Hursley, Winchester, SO21 2JN, United Kingdom}
\author{Freya Shah}
\affiliation{Department of Engineering Science, University of Oxford, Parks Road, Oxford OX1 3PJ, United Kingdom}
\author{Minjun Jeon}
\affiliation{IBM Quantum, T. J. Watson Research Center, Yorktown Heights, NY 10598, USA}
\affiliation{Department of Materials, University of Oxford, Parks Road, Oxford OX1 3PH, United Kingdom}
\author{M. Emre Sahin}
\affiliation{The Hartree Centre, STFC, Sci-Tech Daresbury, Warrington WA4 4AD, UK}
\author{Christa Zoufal}
\affiliation{
IBM Quantum, IBM Research Europe — Zurich, Ruschlikon 8803, Switzerland}
\author{Kunal Sharma}
\email{kunals@ibm.com}
\affiliation{IBM Research, Chicago, IL 60606, USA}
\date{\today}

\begin{abstract}
Recent theoretical progress has established conditions under which machine learning models can efficiently predict ground-state properties of gapped local Hamiltonians when trained on quantum-generated data. Previous experimental demonstrations in this paradigm, however, have largely been limited to small systems or highly structured states, due to the difficulty of preparing many-body ground states on quantum processors. In this work, we demonstrate learning from experimental quantum data generated from approximate ground states of the two-dimensional Heisenberg XXZ model with system sizes up to 115 qubits. We construct a dataset of single-site expectation values, two-point correlations, and 12-body loop correlations across the antiferromagnetic phase. We then train neural networks on this data and show that they can accurately predict spatially resolved observables for previously unseen Hamiltonian parameters, both within the training distribution and in an out-of-distribution regime approaching the phase boundary. Our results demonstrate the practical realization of learning from quantum data for an interacting two-dimensional many-body system at scale, motivating a path toward regimes where quantum processors could provide training data beyond the reach of classical approximation methods. 

\end{abstract}

\maketitle

\newacronym{dmrg}{DMRG}{density matrix renormalization group}
\newacronym{skqd}{SKQD}{sample-based Krylov quantum diagonalization}
\newacronym{afm}{AFM}{antiferromagnetic}
\newacronym{ml}{ML}{machine learning}
\newacronym{peps}{PEPS}{projected entangled pair states}
\newacronym{obp}{OBP}{operator backpropagation}
\newacronym{hpc}{HPC}{high-performance computing}
\newacronym{qpu}{QPU}{quantum processing unit}
\newacronym{mps}{MPS}{matrix product state}
\newacronym{ipr}{IPR}{inverse participation ratio}
\newacronym{cv}{CV}{cross-validation}

\section*{Introduction}
\label{sec:introduction}

\begin{figure*}[ht]
    \centering
    \includegraphics[width=\textwidth]{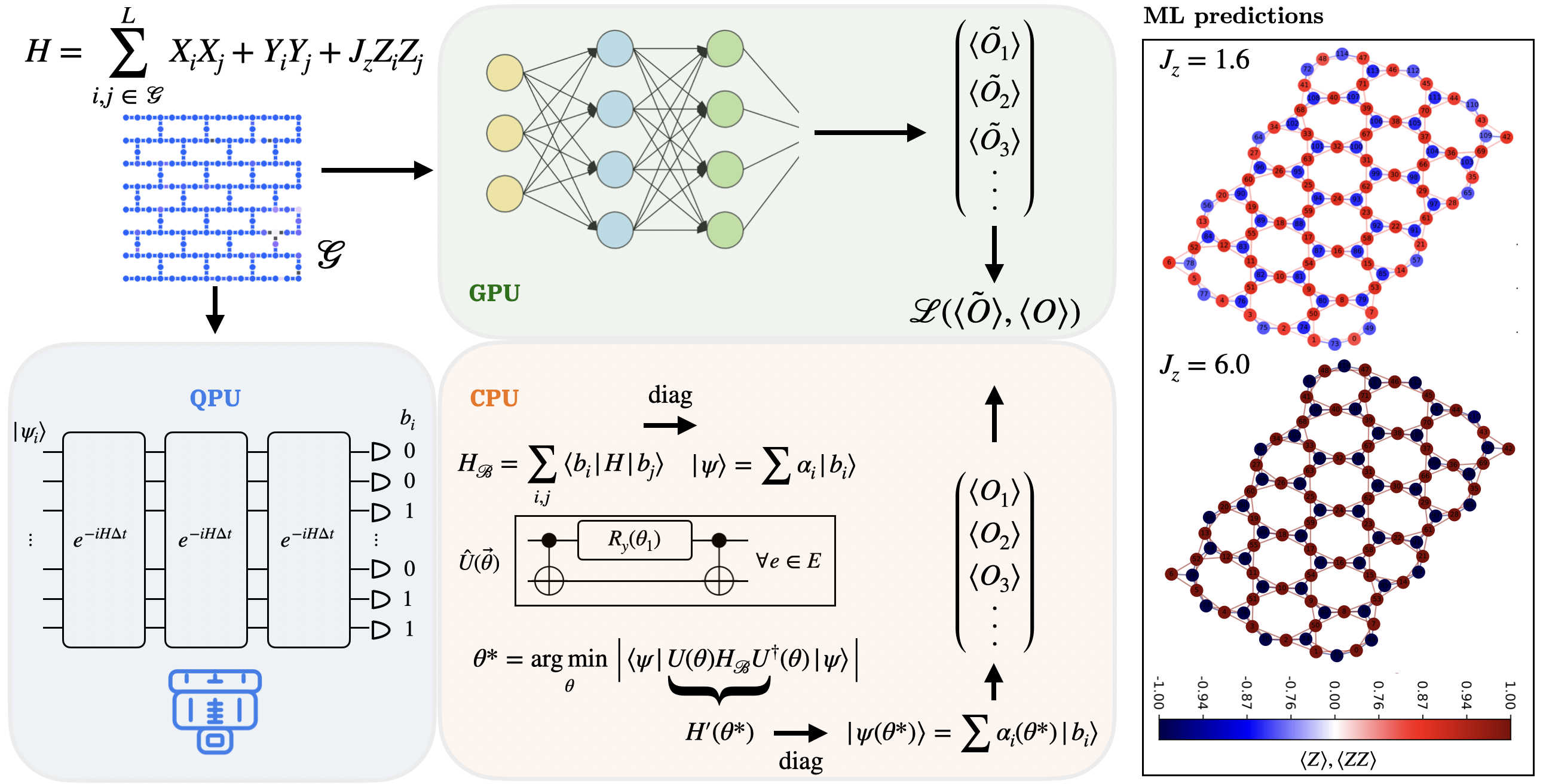}
    \caption{Overview of the workflow. (Left) For a fixed heavy-hex lattice and anisotropy $J_z$ of the spin-$\frac{1}{2}$ Heisenberg XXZ model, we generate approximate ground states by first running sample-based Krylov diagonalization (SKQD) on the \textit{ibm\_boston} quantum processor (blue box). This is followed by a basis-optimization step (orange box) that learns a shallow unitary transformation of the Hamiltonian which further lowers the energy when diagonalized. We generate data by computing local observables from the resulting state, which are used to train a neural network via supervised learning (green box). (Right) Once trained, the model predicts these observables at previously unseen values of $J_z$ across the parameter space.}
    \label{fig:1}
\end{figure*}

Predicting properties of quantum many‑body ground states is central to condensed‑matter physics~\cite{eisert2010colloquium, verstraete2004renormalization, amico2008entanglement}, quantum chemistry~\cite{kohn1965self}, and quantum information science~\cite{latorre2003ground}. Observables such as energies, correlation functions, and order parameters characterize phase structure and material properties of quantum systems. In many instances of practical interest, however, strong interactions and large system sizes render an exact classical treatment intractable. Furthermore, many scientific problems require studying not a single Hamiltonian, but families of Hamiltonians parameterized by couplings, fields, or geometric variations~\cite{vojta2003quantum, arovas2022hubbard, janssen2019heisenberg}. Repeatedly evaluating ground‑state properties across such parameter spaces can incur substantial computational cost, creating a major bottleneck for many-body simulation. 

This has motivated growing interest in \gls{ml} approaches that seek to amortize this cost by training models to approximate the map from Hamiltonian parameters to ground‑state observables using previously generated data~\cite{behler2007generalized, ramakrishnan2015big, carrasquilla2017machine, brockherde2017bypassing, schmidt2017predicting, ch2017machine}. Once trained, such models can be queried cheaply across parameter space, enabling rapid exploration without rerunning the underlying procedure at each point. In the setting where the training data are generated by quantum computers, recent theoretical work has begun to formalize when such learning is possible.

For families of gapped local Hamiltonians, classical learners have been shown to efficiently predict local ground-state properties across parameter ranges that remain within the same quantum phase~\cite{huang2022provably}, given training data obtained from measurement schemes such as classical shadows~\cite{huang2020predicting}. Subsequent results showed that incorporating geometric locality as an inductive bias can substantially reduce the required amount of training data and improve scaling with system size~\cite{lewis2024improved}, while more recent analyses extend these guarantees to broader classes of models, including certain deep learning architectures~\cite{wanner2024predicting}. However, numerical experiments supporting these theoretical developments have largely used \gls{dmrg}~\cite{schollwock2005density}, even though it is not computationally efficient for 2D systems~\cite{stoudenmire2012studying}. A central open question is therefore whether this learning paradigm can be realized directly on large-scale quantum hardware for interacting many-body systems.

Experimental progress towards learning from quantum data has so far been limited by the difficulty of preparing accurate many‑body states and estimating their observables on noisy hardware. Recent experiments have demonstrated proof-of-principle results for systems that admit exact classical solutions and have canonical (fixed-point) states that represent the phase but can be prepared with shallow circuits~\cite{cho2024machine}. Therefore, the practical realization of learning from quantum experiments for larger, interacting many‑body ground states remains an open challenge.

In this work, we address this challenge for ground states of the Heisenberg XXZ model on two-dimensional heavy-hex lattices with up to 115 qubits. Although recent sample-based quantum diagonalization methods have shown promising results~\cite{yu2025quantum}, they do not directly apply across the full parameter regime studied here, because the target ground states are not generally sparse in the computational basis used for sampling. To overcome this limitation, we develop a workflow that combines \gls{skqd}~\cite{yu2025quantum} with a basis-optimization procedure that learns an entangled basis in which diagonalizing the projected Hamiltonian yields a lower-energy approximate ground state. Running for over 8 hours on the \textit{ibm\_boston} superconducting quantum computer, and using \gls{hpc} resources for diagonalization and basis optimization, we construct ground states at 45 points in the \gls{afm} phase for both $n=57$ and $n=115$ qubits. From these states, we compute 1-, 2- and 12-local observables, forming a training dataset with few-percent error compared to \gls{dmrg} across most of the phase.

We use these observables as supervised learning targets to train neural networks that predict ground‑state properties as a function of the Hamiltonian parameters. Figure~\ref{fig:1} summarizes the end-to-end workflow, from quantum data generation to model training and prediction. Training on data for $1.6 \leq J_z \leq 6.0$, we find that the models generalize to unseen Hamiltonian parameters within this regime, accurately predicting local observables across the lattice. Beyond the training distribution and towards the phase boundary at $J_z=1.0$, they continue to do so with percent-level relative error despite changes in the underlying ground-state physics.

At $J_z = 1.1$, the models also capture spatial variations of the observables, including differences between boundary and bulk lattice sites. Overall, our results demonstrate that classical machine learning can leverage data from current quantum computers to predict ground-state observables at unseen Hamiltonian parameters in interacting two-dimensional many-body systems, pointing toward regimes where quantum data could enable learning beyond the reach of classical approximation methods.

\section*{Results}
\label{sec: results}

\subsection{Verifying data quality}

\begin{figure*}[ht]
    \centering
    \includegraphics[width=\linewidth]{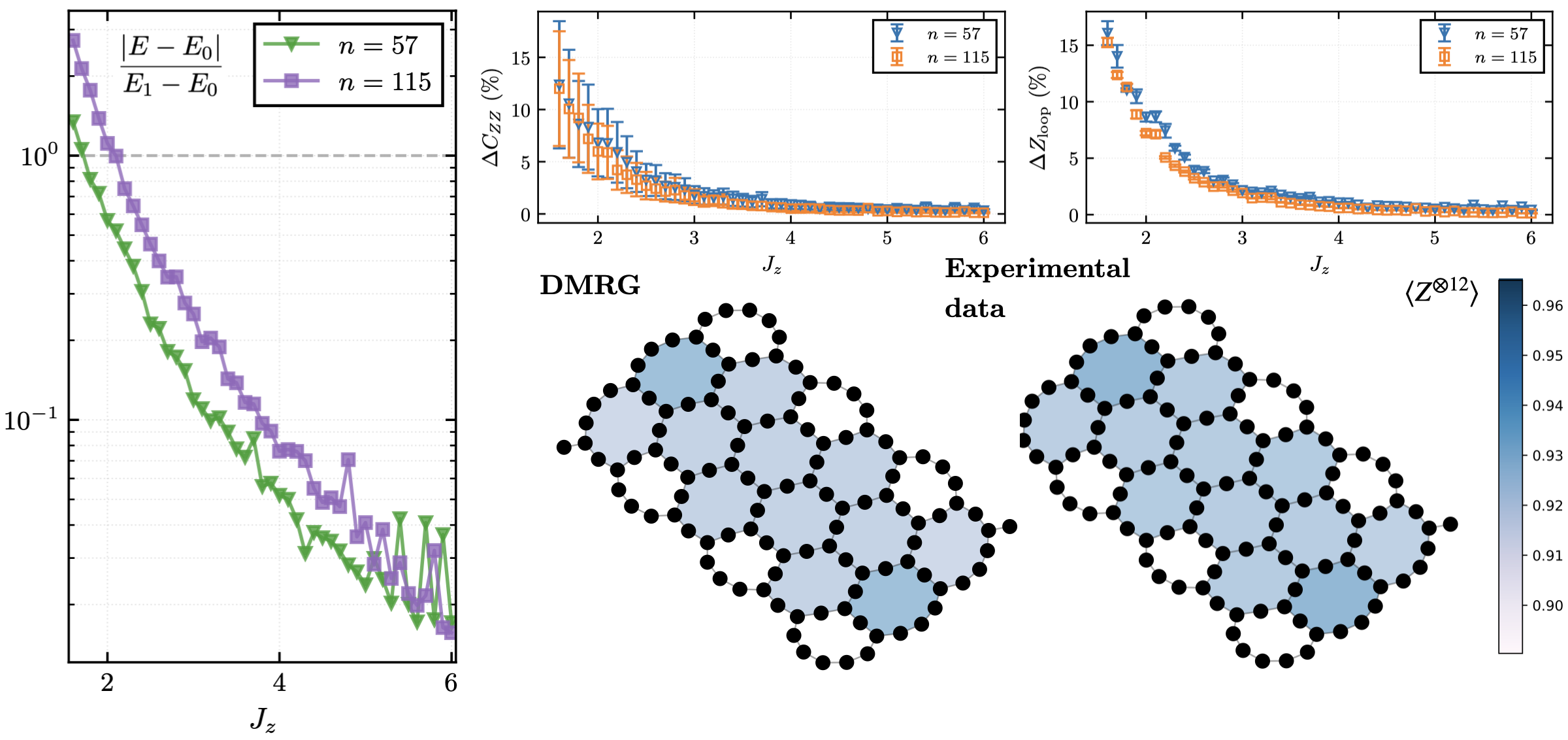}
    \caption{Accuracy of data generated from quantum experiments. (Left) Error in the approximate ground state energy $E$ relative to the gap to the first excited state $(|E-E_0|)/(E_1-E_0)$ as calculated by DMRG. (Top) Mean percentage error between the quantum data and DMRG for $C_{zz}$ and $Z_{\mathrm{loop}}$, error bars show one standard deviation of the distribution of individual relative errors. (Bottom) Spatial map of $\langle Z_{\mathrm{loop}}\rangle$ for $n=115$ at $J_z=4.0$. Colors denote 12-qubit loop values and white regions indicate 10-qubit loops not evaluated.}
    \label{fig:2}
\end{figure*}

We first demonstrate that the approximate ground states produced by our workflow yield local observables that agree closely with a converged classical reference across most of the phase. We study the spin-$\frac{1}{2}$ Heisenberg XXZ Hamiltonian

\begin{equation}
    H(J_z) = \frac{1}{4}\sum_{\langle i,j\rangle \ \in \ \mathcal{G}}X_iX_{j} + Y_iY_{j} + J_{z}Z_iZ_{j},
\end{equation}

where $P_i \in \{X, Y, Z\}$ are Pauli matrices, $\langle i,j\rangle$ denotes nearest neighbors on a two-dimensional heavy-hexagonal~\cite{chamberland2020topological} graph $\mathcal{G}$ and $J_z$ tunes the anisotropy of the model. In this work, we consider graphs of distance 5 and 7, corresponding to $n=57$ and $n=115$ qubits, respectively. The heavy-hexagonal geometry is chosen because it matches the native connectivity of the superconducting processor, while retaining non-trivial two-dimensional Hamiltonian interactions.

We generate approximate ground states of the Hamiltonian in the \gls{afm} phase using our ground-state workflow that begins with \gls{skqd}~\cite{yu2025quantum, kirby2026observation}. This method constructs a low-dimensional Krylov subspace by sampling from time-evolution circuits on quantum hardware and then solves the projected eigenvalue problem classically to obtain an approximate ground state. A key assumption underlying this algorithm is that the target state is sparse in the sampled basis. For the 2D XXZ model studied here, however, this assumption is nontrivial because the ground states across the phase are not generally sparse in the computational basis. We therefore introduce a variational basis optimization step, which identifies an entangled Hamiltonian basis in which the ground state energy is lower when projected into the sampled subspace. This basis transformation is defined by a fixed-depth unitary whose action on the Hamiltonian can be efficiently simulated classically using \gls{obp}~\cite{fuller2026improved}, based on Pauli propagation~\cite{rall2019simulation}. The memory requirements of these steps for large subspaces necessitate \gls{hpc} resources. Experimental details for \gls{skqd} and an overview of the basis-optimization procedure are provided in \hyperref[sec:methods]{Methods}.

As a reference benchmark, we compare against \gls{dmrg} calculations with bond dimension $\chi=320$, performed using the TeNPy library~\cite{tenpy2024}. The \gls{dmrg} results are used only for benchmarking and not used as supervised labels for the neural networks. While \gls{dmrg} is itself approximate in 2D, it is a standard and well-controlled method at these system sizes. In Supplementary~\ref{supp:sec:dmrg_convergence} we show convergence of the energy and local observables for the instances considered here. Figure~\ref{fig:2} summarizes the resulting agreement with \gls{dmrg}. In the left panel, we plot the difference between the ground state energy produced by our algorithm ($E$) and DMRG ($E_0$), respectively, expressed in units of the many-body gap $E_1 - E_0$. This quantity decreases rapidly as $J_z$ increases. For $n=57$, the experimental energies lie below the gap for $J_z \geq 1.8$, while for $n=115$ this occurs for $J_z\geq2.1$. Whether the approximate ground-state energy lies below the spectral gap is a useful benchmark because it places an upper bound on the distance to the true ground state, as explored further in Supplementary~\ref{supp:sec:skqd_vd_dmrg}. We note that data generation towards the phase boundary is inherently difficult for both classical and quantum methods. 

The approximate ground states obtained from our workflow admit a classical representation and thus could be used directly as training data for \gls{ml} models. However, we use expectation values of selected observables as the training data, which can be evaluated efficiently from the approximate ground state. These observables are the single-site magnetization $\langle Z \rangle$, the two-point correlator $C_{zz} = \langle Z_i Z_j \rangle$ over first- and second-nearest-neighbor edges, and the loop correlator 

\begin{equation}
   Z_{\mathrm{loop}} = \left\langle \prod_{k \in \ell_{12}} Z_k \right\rangle, 
   \label{eqn:z_loop}
\end{equation}

where $\ell_{12}$ denotes a fixed loop of twelve sites on the lattice. In the top panel of Fig.~\ref{fig:2}, we quantify the accuracy of $C_{zz}$ and $Z_{\mathrm{loop}}$. Here we plot the mean relative error between \gls{dmrg} and the quantum data, which is defined for an observable family $O$ as

\begin{equation}
    \Delta O_{\textrm{data, DMRG}}(\%) = \frac1K\sum_k^K \frac{|\langle O_k\rangle_{\mathrm{data}} - \langle O_k\rangle_{\mathrm{DMRG}}|}{\langle O_k\rangle_{\mathrm{DMRG}}}.
\label{eqn:average_relative_error}
\end{equation}

Here $K$ is the number of observables in that family over the lattice, for example $K=n$ for $\langle Z \rangle$. In both cases, the error increases as $J_z$ decreases, following the same pattern observed in the energy. Across the phase space, the median relative errors for $n=57$ and $n=115$ are $M(\Delta C_{zz})$ = 1.36\% and 1.22\%, $M(\Delta Z_{\mathrm{loop}})$ = 1.57\% and 1.27\%, respectively. Notably, the loop correlators achieve accuracy comparable to the two-point correlations despite being higher-weight observables. This shows that our workflow yields quantum data that accurately captures not only the energy but also nontrivial multi-qubit structure, establishing a large-scale experimental benchmark for the 2D Heisenberg XXZ model. Further analysis, including the error for the single-site magnetization $\langle Z \rangle$, is given in Supplementary~\ref{supp:sec:skqd_vd_dmrg}.

Finally, the bottom panel of Fig.~\ref{fig:2} shows that the loop correlator data preserve the correct spatial variations across the lattice. At $J_z=4.0$ for $n=115$, the \gls{dmrg} loop pattern exhibits two boundary-adjacent loops with larger values than those in the bulk. The approximate ground state produced by our workflow reproduces this pattern despite the high precision required to resolve it. This shows that the constructed dataset preserves local structure rather than only matching spatial averages, motivating its use as training targets for learning ground-state properties.

\subsection{In-distribution generalization}

\begin{figure}[ht]
    \centering
    \includegraphics[width=\linewidth]{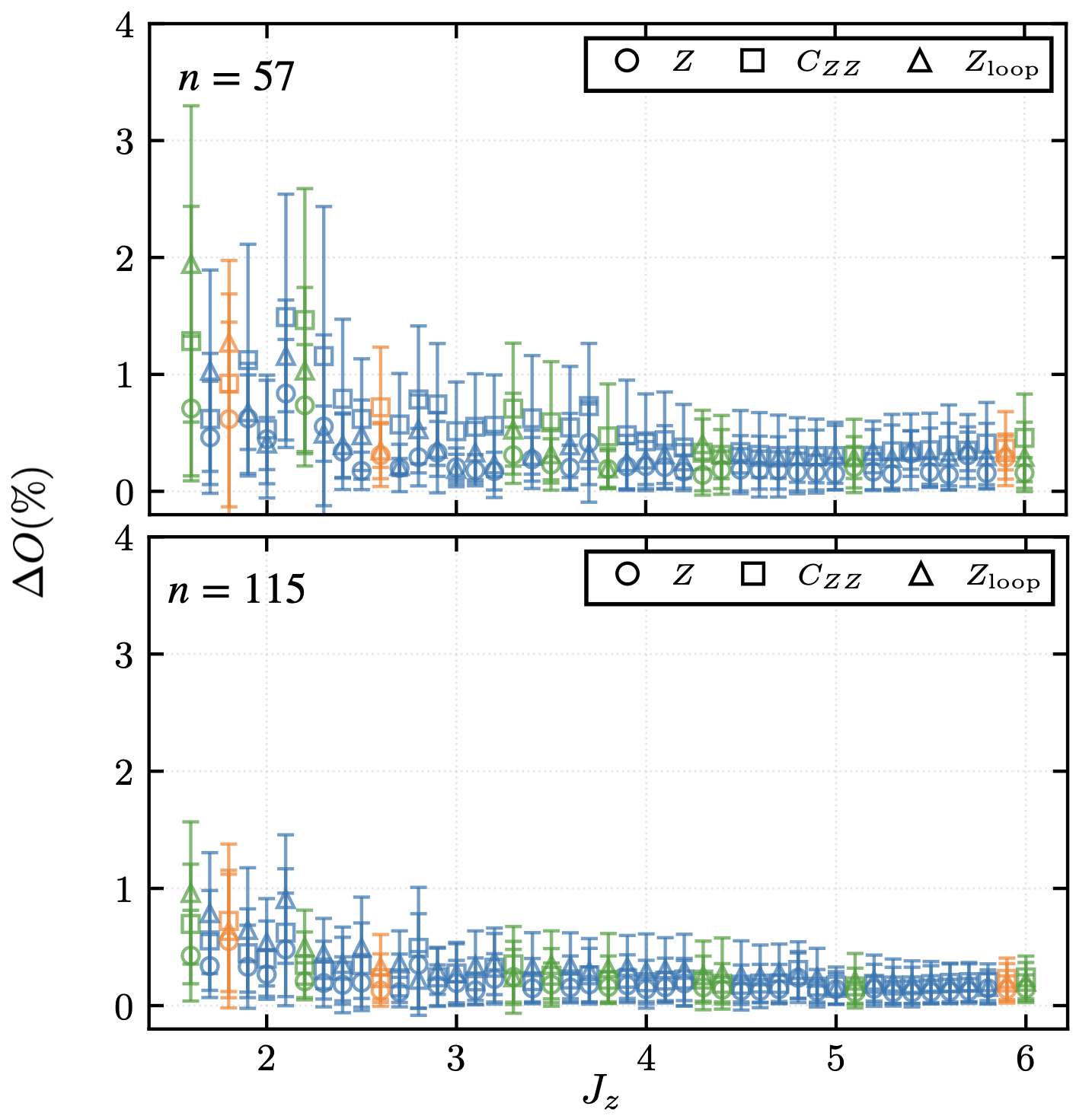}
    \caption{Model generalization within the interior of the AFM phase as shown by mean relative error $\Delta O_{\textrm{ML, data}}(\%)$. Error bars show one standard deviation of the distribution of individual relative errors. Blue, orange and green markers correspond to $J_z$ points from the train, validation and test data set respectively.}
    \label{fig:3}
\end{figure}

We next show that neural network models trained on this data can generalize to accurately predict local observables at unseen points within the interior of the \gls{afm} phase, providing a practical realization of the in-phase learning setting analyzed theoretically in Refs.~\cite{huang2022provably,lewis2024improved,wanner2024predicting}. We use the architecture introduced in Ref. ~\cite{wanner2024predicting} which combines the outputs of several local multi-layer perceptrons to predict a single observable $\langle \tilde{O}_1\rangle$. These local predictors are then stacked to create a model whose output is the vector of observables evaluated in the supervised loss function. Further details of the architecture, together with a comparison to a simpler architecture without geometric locality, are provided in Supplementary~\ref{supp:sec:deep_learning}.

To evaluate generalization within the phase, we use training data from the parameter range $1.6 \leq J_z \leq 6.0$ and partition this interval into three strata: $4.1$--$6.0$, $2.1$--$4.0$, and $1.6$--$2.0$. These strata correspond to ground states with progressively smaller spectral gaps. Since ground-state preparation generally becomes more difficult as the spectral gap decreases, we expect smaller gaps to reduce state-preparation fidelity for a fixed resource budget. Consequently, we expect increased error in this regime, both for our workflow and for ground-state preparation methods more broadly~\cite{lin2020near}. Thus, we consider it important for each region to have equal relative proportion in training and testing. We construct a held-out test set by removing $20\%$ of points from each stratum. From the remaining data, we select hyperparameters for each model by cross-validation over training and validation splits, and then perform a final retraining using the selected hyperparameters. Full details are given in Supplementary~\ref{supp:sec:deep_learning}.

We train three separate models to predict expectation values across the lattice for the observable families $Z$, $C_{zz}$ and Eq.(\ref{eqn:z_loop}), respectively. The corresponding number of observables for $n=57$ ($n=115$), and hence the output dimension of each model, is 57 (115), 151 (317) and 4 (12). Although the Hamiltonian is parameterized here by a single scalar $J_z$, the learning task remains nontrivial because each input must be mapped to a spatially resolved set of many observables across the lattice, including higher-locality correlations and non-uniform boundary and bulk structure.

Figure~\ref{fig:3} summarizes the performance of trained models. The plotted quantity is the average relative error between the \gls{ml} predictions and quantum data, defined analogously to Eq.(\ref{eqn:average_relative_error}) and denoted here by $\Delta O_{\textrm{ML, data}}(\%)$. Here error bars show one standard deviation of the spread of relative error across instances of each observable at fixed $J_z$. For $n=57$, the models achieve $\le 2\%$ error for all three observable types, with a slight increase near the lower end of the interval where the ground-state properties vary more rapidly with $J_z$. The held-out test points in orange follow the same trend as the training and validation points in blue and green respectively, indicating reliable generalization within the phase. 

For $n=115$, we observe a similar trend, but with a lower average error, remaining below $1\%$ across all observable families and points in the phase space. Overall, Fig.~\ref{fig:3} shows that within the interior of the \gls{afm} phase, models trained on a limited set of quantum data points can predict spatially resolved local observables at unseen $J_z$ with percent-level accuracy on both system sizes.

\begin{figure*}[ht]
    \centering
    \includegraphics[width=\linewidth]{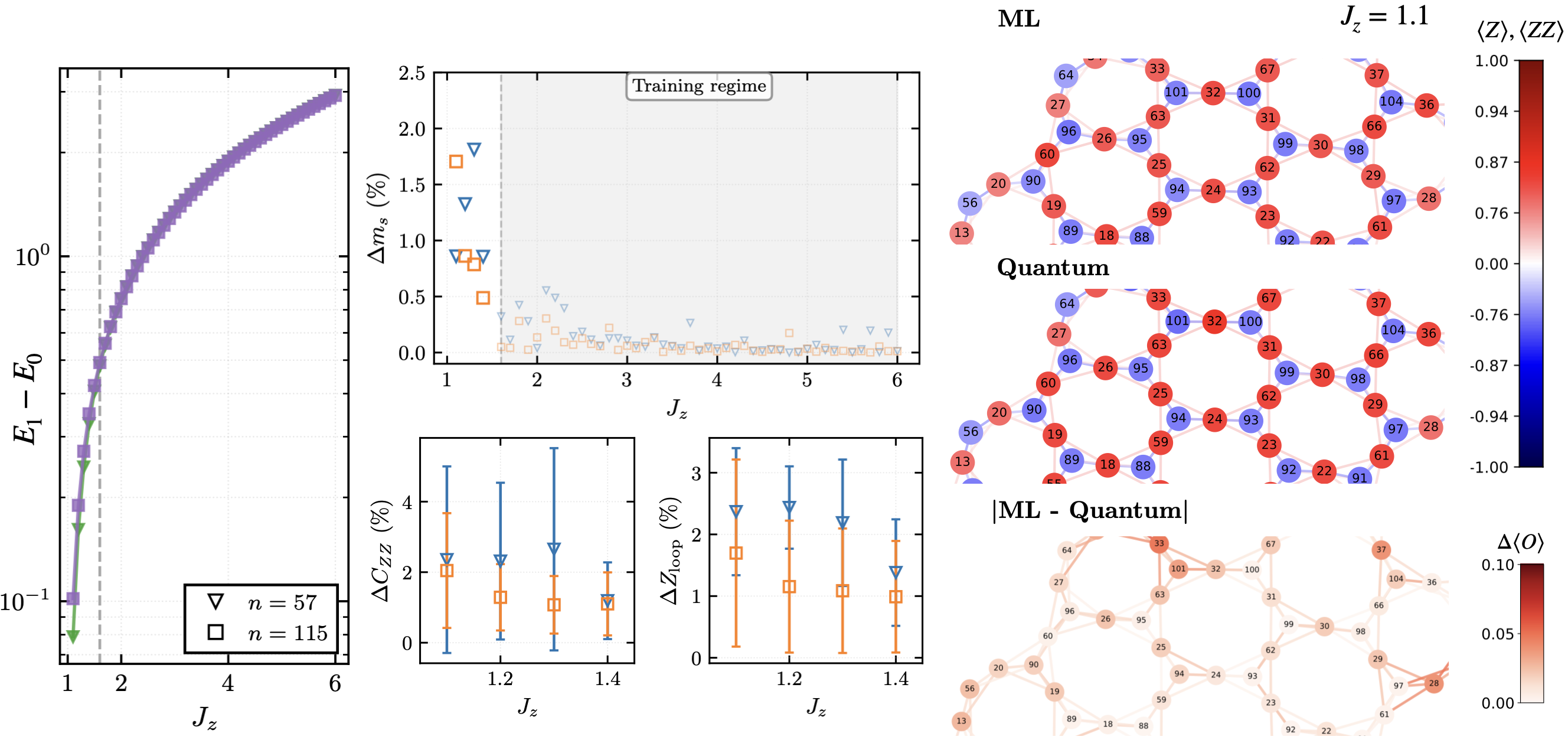}
    \caption{Model generalization towards the phase boundary. (Left) Spectral gap $E_1-E_0$ as a function of $J_z$ on approaching the phase boundary at $J_z=1.0$, which rapidly decreases below the training cutoff at $J_z = 1.6$. (Middle) Relative error between local observables predicted by \gls{ml} and data from new quantum experiments outside the training regime. This includes the staggered magnetization $m_s$, an order parameter of the AFM phase which is inferred from the model trained on $\langle Z \rangle$. (Right) Representative slice through the lattice at $J_z=1.1$, comparing the spatially resolved observables from \gls{ml} and the quantum experiments, together with the absolute difference.}
    \label{fig:4}
\end{figure*}

\subsection{Out-of-distribution generalization}

We next evaluate the ability of the models to generalize outside of the training interval $1.6 \leq J_z \leq 6.0$ and towards the phase boundary at $J_z=1.0$. We train a new set of models on all data in the range $1.6 \leq J_z \leq 6.0$, including points previously held out for the in-distribution test. We use the same hyperparameters as before and define a small validation set of 5\% of the points in each stratum.

In general, computing ground states near a phase boundary is difficult for both classical and quantum methods. For the XXZ model, the transition at $J_z=1.0$ from the gapped \gls{afm} phase to the gapless critical XY phase~\cite{lee2023quantum, kosterlitz1973ordering} is marked by the closing of the spectral gap. The left panel of Fig.~\ref{fig:4} shows the many-body gap, computed by \gls{dmrg}, as a function of $J_z$. The gap decreases rapidly as $J_z$ approaches the phase boundary, with the most pronounced change occurring below the training cutoff at $J_z = 1.6$. This identifies the region $1.1 \leq J_z \leq 1.4$ as one in which the ground-state physics changes over a short parameter interval, making it an important testbed for the generalization capabilities of the trained models.

The middle panels quantify how accurately the models predict observables produced by the quantum workflow in this regime. To do this, we run an additional set of experiments for $1.1 \leq J_z \leq 1.4$ to generate test data, see \hyperref[sec:methods]{Methods}. Subsequently, we consider the relative error between the \gls{ml} prediction and quantum experiments for the staggered magnetization $m_s = \frac{1}{N}\sum_i \eta_i \langle Z_i \rangle$, where $\eta_i \in \{\pm 1\}$ assigns a sign according to the antiferromagnetic sub-lattice pattern. This observable is not explicitly learned but is derived from the model trained to predict $\langle Z \rangle$ at each site. Despite the absence of training data below $J_z = 1.6$, the models achieve $<3\%$ error in this region across both system sizes, albeit with a clear increase in error. Similarly, for the two-point correlators $C_{zz}$ and the loop correlators, the spatially averaged error remains below $4\%$ across the extrapolation region. This shows that the models remain accurate when extrapolating toward the phase boundary, where the closing spectral gap makes the ground-state observables vary more rapidly.

Finally, the right panel shows that the models capture spatial variations of the observables at $J_z=1.1$, the closest point to the phase boundary considered here. For a representative slice through the lattice, the \gls{ml} predictions reproduce the structure of the new quantum data across the lattice, including differences between boundary (e.g., qubit 20) and bulk (e.g., qubit 25). Taken together, Fig.~\ref{fig:4} shows that the models generalize beyond the training distribution not only in terms of averaged observables, but also in their specific spatial structure.

\section*{Discussion}

In this work, we demonstrate an end-to-end workflow for learning ground-state observables from experimental quantum data in a large, two-dimensional interacting many-body system. Combining \gls{skqd} with an optimized entangled-basis transformation of the projected Hamiltonian, we obtain approximate ground states of the Heisenberg XXZ model on heavy-hex lattices with up to 115 qubits. From these states, we compute a dataset of 1-, 2- and 12-local observables that agrees with \gls{dmrg} at the percent level across a broad region of the antiferromagnetic phase. This establishes a practical setting in which \gls{ml} models can be trained and tested on experimental data from quantum computers at system sizes beyond the reach of exact classical simulation.

Once trained on this dataset, the neural networks accurately predict ground-state observables at unseen Hamiltonian parameters. Within the training interval $1.6 \le J_z \le 6.0$, the models generalize with small relative error for both system sizes considered. Beyond the training distribution, near the phase boundary, they continue to reproduce the corresponding observables from new quantum experiments with high accuracy, including the spatial structure of the data at the closest point to the phase boundary considered here. Understanding more systematically the generalization properties of \gls{ml} models trained on quantum ground-state data is an important direction for future work, particularly their dependence on system size, observable locality, and proximity to phase boundaries.

Another natural extension of our work is to move beyond learning selected observables and instead learn compact classical representations of quantum-computed ground states themselves, for example, from sample-based methods used in this work. This mirrors aspects of neural quantum states, but in a supervised setting where quantum computation provides a training signal rather than relying only on variational optimization~\cite{carleo2017solving}. Once such a representation is learned, it could support efficient evaluation of a broader class of observables than those specified during training, including high-locality operators such as the loop correlators considered here. More generally, this points to a route in which quantum experiments provide rich ground-state data that classical learning models can leverage for broader downstream analysis across families of Hamiltonians.

Looking further ahead, the broader motivation for this workflow is the prospect of using quantum processors to generate training data for states that are increasingly difficult to access with classical methods. Although ground states of gapped two-dimensional local Hamiltonians are widely expected to satisfy an entanglement area law, and this has been proven for important subclasses~\cite{eisert2010colloquium, anshu2022area}, such scaling alone does not imply the existence of an efficient tensor network representation with \gls{peps}~\cite{verstraete2004renormalization, ge2016area}. Together with the known hardness of contracting \gls{peps}~\cite{schuch2007computational}, quantum computing offers a scalable alternative to generating ground state data. More broadly, as quantum hardware improves, such workflows could enable classical models to learn from quantum-generated data in regimes where classical approximation methods become prohibitively expensive. 

\section*{Methods}
\label{sec:methods}

\subsection*{Sample-based Krylov quantum diagonalization experimental details}
\label{methods:experimental_details}

In \gls{skqd}, a basis set is constructed in a subspace smaller than the full Hamiltonian by sampling from circuits implementing the time-evolution of the Hamiltonian. Here the key ingredients are the initial state $|\psi_i\rangle$ to begin time evolution from, the Krylov dimension $k$ corresponding to the maximum number of time steps taken and $\Delta t$ the time step size. For further details on \gls{skqd} we refer to Refs.~\cite{yu2025quantum, kirby2026observation, piccinelli2025quantum}.

We choose $|\psi_i\rangle$ to be the N\'eel state, a product state of alternating up and down spins, and is the ground state of the system in the limit $J_z \rightarrow \infty$. Since the rate of convergence for \gls{skqd} depends on the overlap between this initial state and the true ground state, this represents a sensible choice. Furthermore it can be implemented with only one layer of single-qubit $X$ gates. 

We choose Krylov dimension $k=10$, meaning that for each $J_z$ point and number of qubits $n$, we evaluate 10 different circuits corresponding to applying 1 to 10 Trotter steps to $|\psi_i\rangle$. The final circuits have a $CZ$-depth of 91 and 3960 (1920) $CZ$ gates for $n=115$ ($n=57$). This represents a balance: choosing a larger $k$ can increase the quality of the subspace, but costs more quantum resources and the deeper circuits generate more noise which affects the likelihood to sample from the true unitary evolution. We did not perform extensive testing of different $k$ values and our ground-state data may be further improved by doing so. 

We choose $\Delta t$ differently for each $J_z$ using a Bayesian optimization scheme from low bond dimension belief propagation simulation of the circuits, see Supplementary~\ref{supp:sec:skqd}.

All \gls{skqd} circuits were evaluated on the \textit{ibm\_boston} quantum computer, with $10^5$ shots sampled from each of the 10 Krylov circuits which were submitted together as one ``job''. The Qiskit \texttt{SamplerV2} primitive~\cite{javadi2024quantum} was used with dynamical decoupling~\cite{pokharel2018demonstration} and Pauli twirling~\cite{wallman2016noise} error mitigation switched on. Each job took approximately 5 minutes of \gls{qpu} time and corresponds to one $J_z$ data point. Thus to generate the data set $J_z=1.6$ to $J_z=6.0$ in increments of 0.1, as well as an extrapolation test region $J_z=1.1$ to $J_z = 1.4$, 50 jobs were submitted for $n=57$ and $n=115$ respectively, for a total quantum resource overhead of 100 million shots over 500 minutes (8.3 hours) of usage. 

A single job ($n=115$, $J_Z=3.1$) was resubmitted due to erroneously high energy, bringing the total overhead for the data set to 505 minutes and 101 million shots.

From the 1 million shots per $J_z$, an average of $972759 \pm 10610$ ($999531\pm471$) unique bitstrings were observed for $n=57$ ($n=115$), where here $\pm$ is a range of one standard deviation. Of these, an average of $81862\pm8114$ ($81974\pm8138)$ were in the correct antiferromagnetic sector and automatically included in the Krylov subspace. We applied a configuration recovery scheme based on random bit flipping to the remaining samples, as detailed in Supplementary~\ref{supp:sec:skqd}, resulting in a subspace dimension $\mathcal{D}$ equal to the number of unique bitstrings originally sampled.

The resulting matrix of dimension $n \times \mathcal{D}$ was diagonalized on an \gls{hpc} cluster node with 250GB of memory.

\subsection*{Basis optimization}
\label{methods:basis_optimization}

In the first step of our workflow, \gls{skqd} returns a set of sampled bitstrings $\mathcal{B}=\{\lvert b_j\rangle\}_{j=1}^{L}$, defining a truncated subspace $\mathcal{S}=\mathrm{span}(\mathcal{B})$. As shown in Fig.~\ref{fig:1}, we subsequently project the Hamiltonian into this subspace, $(H_{\mathcal{B}})_{jk}=\langle b_j\lvert H\rvert b_k\rangle$, and diagonalize $H_{\mathcal{B}}$ to obtain an approximate ground state $\lvert \tilde{\psi}\rangle\in\mathcal{S}$.

We sample in the computational basis states, however, it is not known a priori which basis the ground state may be sparse in. If the true ground state has substantial weight on configurations that are unlikely to be observed in this basis then the \gls{skqd} solution may produce poor results. Whilst in other works an analytically known basis transformation is applied~\cite{yu2025quantum}, in this work we perform a variational optimization that finds a unitary transform which rotates the Hamiltonian into an entangled basis which minimizes the energy
$$
H'(\boldsymbol{\theta}) = U(\boldsymbol{\theta})\,H\,U^\dagger(\boldsymbol{\theta}),
\qquad
C(\boldsymbol{\theta})=\langle \tilde{\psi}\lvert H'(\boldsymbol{\theta})\rvert \tilde{\psi}\rangle.
$$
We minimize $C(\boldsymbol{\theta})$ using a parameterized ansatz $U(\boldsymbol{\theta})$ consisting of a CNOT, $RY(\theta)$, CNOT two-qubit block tiled across the heavy-hex in the optimal 3-coloring. $C(\boldsymbol{\theta})$ is evaluated classically using~\gls{obp}.

Finally, at the optimum $\boldsymbol{\theta}^\star$, we re-project and re-diagonalize within the same sampled subspace by diagonalizing $H'_{\mathcal{B}}(\boldsymbol{\theta}^\star)$, where $(H'_{\mathcal{B}})_{jk}=\langle b_j\lvert H'(\boldsymbol{\theta}^\star)\rvert b_k\rangle$, and take the lowest-eigenvalue eigenvector as the basis-optimized state used to compute observables. Specific circuit settings and optimization details are given in Supplementary~\ref{supp:subsec:basis_optimization}.

Basis optimization for each data point was performed on an \gls{hpc} node with 250GB of memory.

\section*{Code availability}
The code used in this work is available upon reasonable request.

\section*{Data availability}
The data that support the findings of this study are available from the
corresponding author upon request.

\section*{Acknowledgements}
We would like to thank Jason Crain, Natalia Ares, and Pranav Vaidhyanathan for their advice and technical discussions.

This work was supported by the Hartree National Centre for Digital Innovation, a collaboration between the Science and Technology Facilities Council and IBM. 

The authors acknowledge the IBM Research CCC cluster for providing resources that have contributed to the research results reported within this paper. The authors would like to acknowledge the use of the University of Oxford Advanced Research Computing (ARC) facility in carrying out this work.

\onecolumngrid

\pagebreak

% \appendix

\vspace{0.5in}

\begin{center}
	{\Large \bf Supplemental Material for {\it Learning ground state observables from quantum computing experiments}}
\end{center}

\section{DMRG convergence}
\label{supp:sec:dmrg_convergence}

In this work, \gls{dmrg}~\cite{schollwock2005density} calculations were performed using the TeNPy library~\cite{tenpy2024}.

\subsection{Energy}

\begin{figure}[ht]
    \centering
    \includegraphics[width=\linewidth]{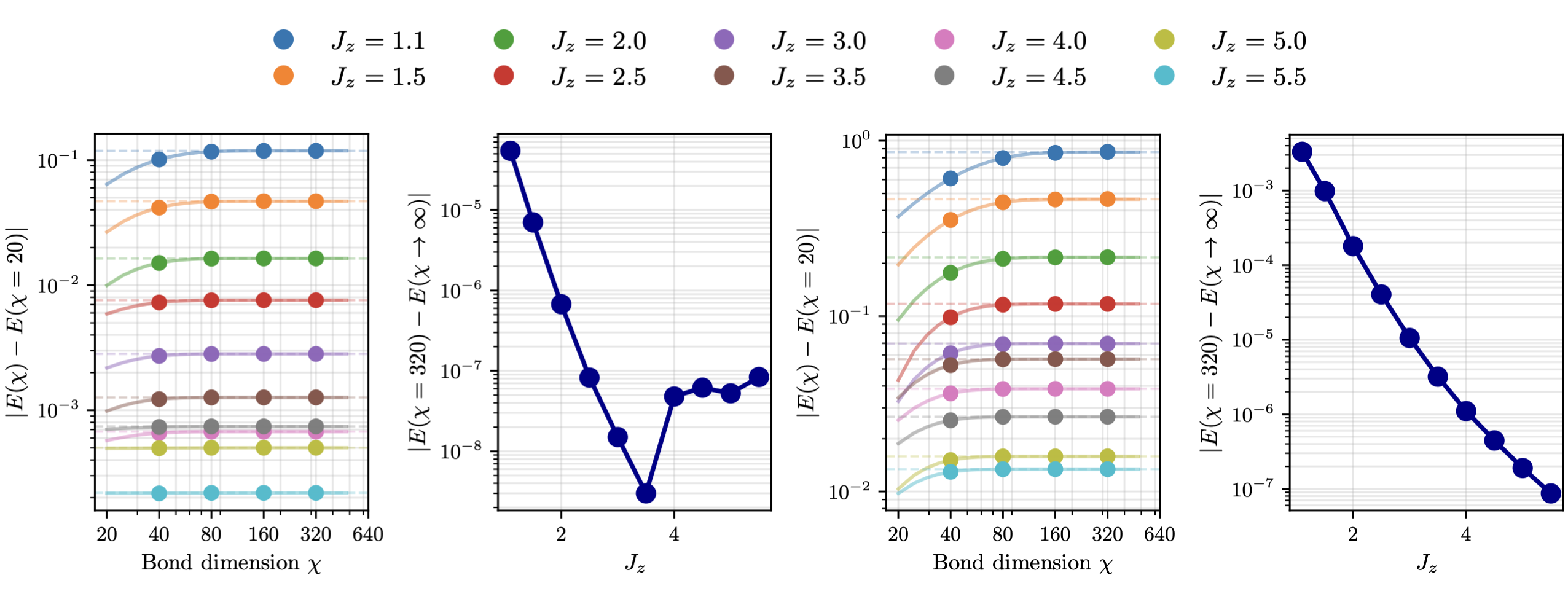}
    \caption{DMRG energy convergence. (Left) For $n=57$, the first plot shows how the energy changes when increasing from $\chi=20$. Circles show specific points evaluated and the solid line shows the trend following an exponential decay fitted to the energies. The offset of the exponential is the energy extrapolated to $\chi\rightarrow\infty$. The second plot shows the absolute difference between $E(\chi=320)$ and $E(\chi\rightarrow\infty)$. (Right) same but for $n=115$.}
    \label{supp:fig:1}
\end{figure}

Fig.~\ref{supp:fig:1} shows the convergence of ground state energy with respect to increasing bond dimension $\chi$, with $n=57$ on the left two plots and $n=115$ on the right two plots. For each $J_z$ we run DMRG with $\chi=[20, 40 ,80, 160, 320]$. The colored plot shows how the predicted energy saturates with increasing $\chi$. Visually we can see that points further into the \gls{afm} phase saturate even for smaller bond dimension, whilst points closer to the phase boundary at $J_z=1.0$ require larger $\chi$. In this work we use $\chi=320$ as our ground truth for all points. Quantifying the error in this choice, we fit the energies to an exponential decay and extrapolate the fit out to $\chi\rightarrow\infty$. The second plot shows the absolute difference in energy between the two values.

\subsection{Observables}

\begin{figure}[ht]
    \centering
    \includegraphics[width=\linewidth]{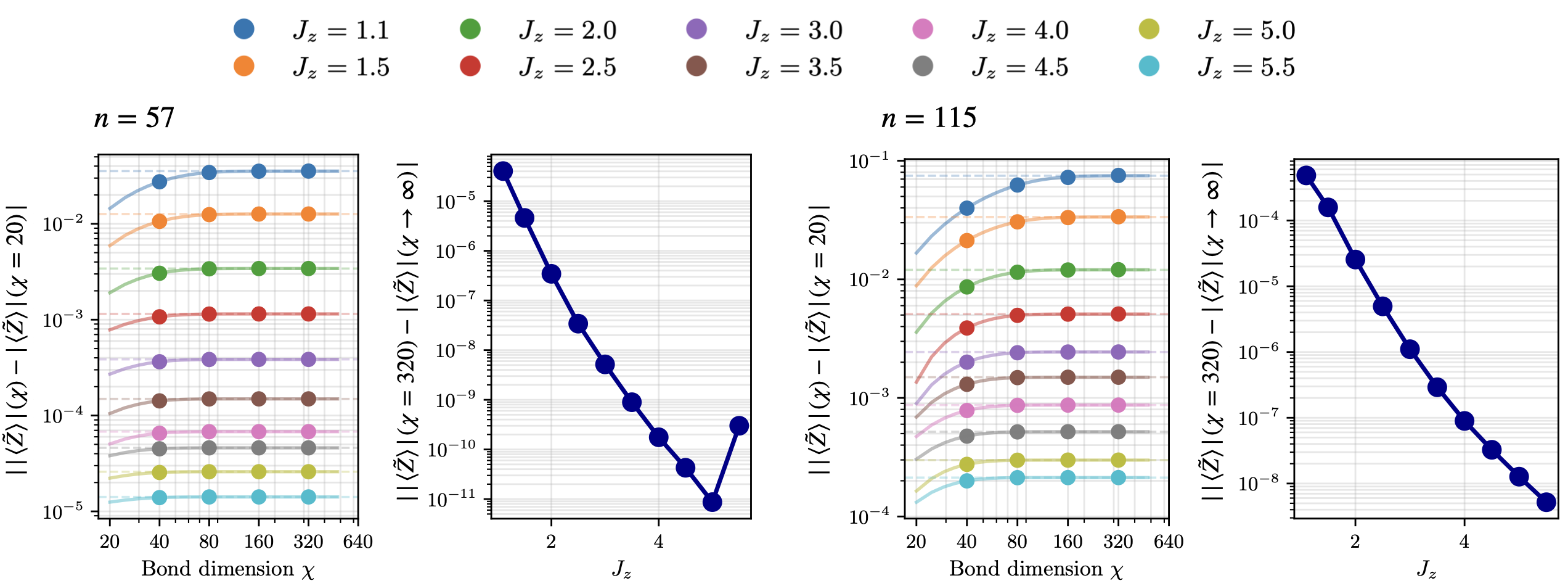}
    \caption{DMRG convergence for the $Z$ observable. (Left) For $n=57$, the first plot shots how the mean absolute expectation value $|\langle \tilde{Z} \rangle| = \frac{1}{L}\sum_i^L|\langle Z_i\rangle|$ changes when increasing from $\chi=20$. Circles show specific points evaluated and the solid line shows the trend following an exponential decay fitted to the average expectation values. The offset of the exponential is the average expectation value extrapolated to $\chi\rightarrow\infty$. The second plot shows the absolute difference between $\langle \tilde{Z} \rangle(\chi=320)$ and $\langle \tilde{Z}(\chi\rightarrow\infty)$.}
    \label{supp:fig:2}
\end{figure}

\begin{figure}[ht]
    \centering
    \includegraphics[width=\linewidth]{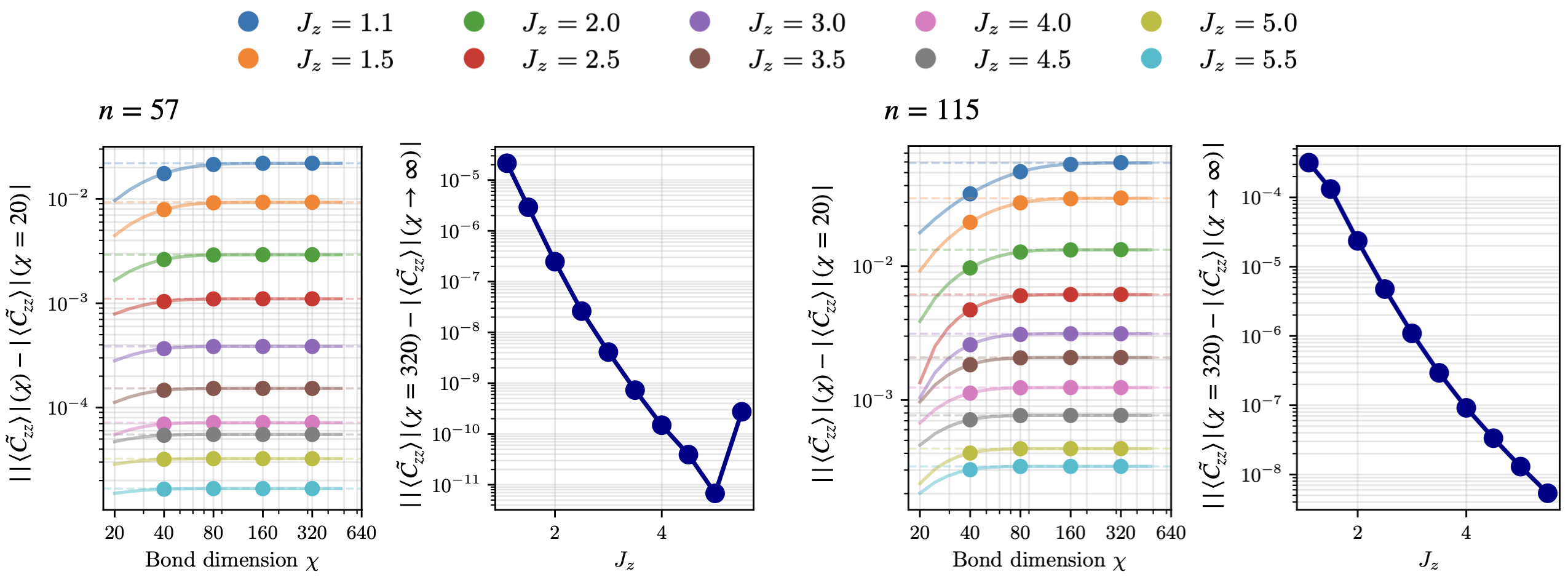}
    \caption{DMRG convergence for the $C_{zz}$ observable. See caption of Fig.2.}
    \label{supp:fig:3}
\end{figure}

\begin{figure}[ht]
    \centering
    \includegraphics[width=\linewidth]{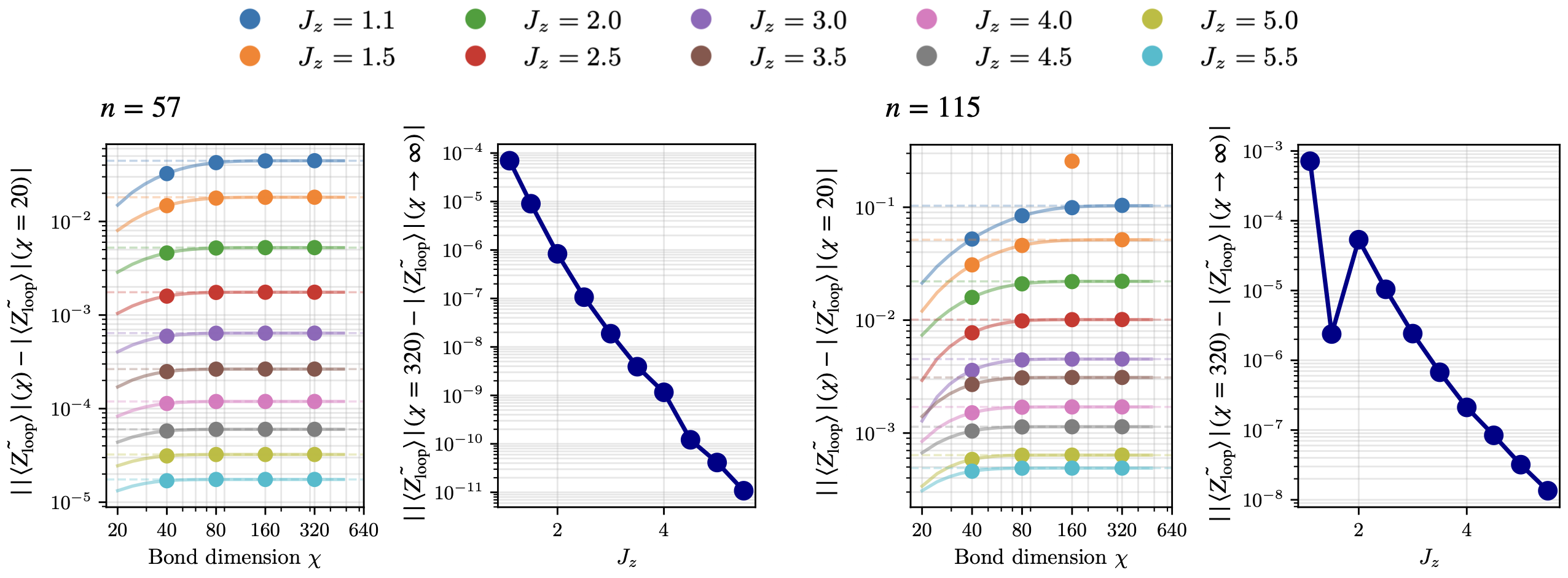}
    \caption{DMRG convergence for the $Z_\mathrm{loop}$ observable. See caption of Fig.2.}
    \label{supp:fig:4}
\end{figure}

We perform a similar analysis to the energy convergence for the expectation values of the single-site magnetization $Z$, 2-site correlator over 1st and 2nd nearest neighbors $C_{zz}$ and the 12-site loop correlator $Z_{\mathrm{loop}}$. As shown in Fig.~\ref{supp:fig:2}, Fig.~\ref{supp:fig:3} and Fig.~\ref{supp:fig:4} respectively, the trends are similar to those seen for the energy and the $\chi=320$ ground state represents a good representation of the state in the limit $\chi\rightarrow\infty$. One outlier occurs in Fig.~\ref{supp:fig:4}, where the loop correlator at $J_z=1.5$ and $\chi=160$ suffers from a numerical error and is not included in the fitting.

\section{Quantum ground-state workflow}
\label{supp:sec:skqd}

\subsection{SKQD $\Delta t$ optimization}

The time step $\Delta t$ considered in the proof of convergence scaling in the original \gls{skqd} work is $\frac{\pi}{|H|}$~\cite{yu2025quantum}. However, practical heuristics suggest that larger time steps around $\frac{25\pi}{|H|}$ can be more resource efficient and give better results for limited $k$~\cite{kirby2026observation}.

To avoid trial-and-error, we construct a Bayesian optimization scheme~\cite{frazier2018tutorial} to optimize $\Delta t$ with the goal of minimizing the energy when running \gls{skqd} with that time step. Here, 3 initial points are randomly sampled from the search space, followed by 7 iterations of Bayesian optimization driven by the acquisition function. This requires 10 \gls{skqd} runs per $J_z$ which we estimate would take $>80$ hours of quantum computing runtime (see Methods of main text). To alleviate this cost, we perform tensor network simulation using the belief propagation method~\cite{tindall2023gauging} and the \texttt{qiskit-tnqs} package~\cite{qiskit-tnqs}. We run the simulations with a bond dimension of $\chi=8$. The simulation incurs both truncation error and error due to the belief propagation ansatz, which we do not rigorously analyze. However, we note that the purpose of this simulation is to approximately find the correct scale of $\Delta t$ and as such our routine does not rely on high bond dimension or our Hamiltonian being inherently classically simulatable to high fidelity.

The optimal $\Delta t$ for each lattice size are shown in Table~\ref{supp:tab:dts}.

\begin{table}[h]
\centering
\setlength{\tabcolsep}{4pt}
\begin{tabular}{|c|c|c||c|c|c||c|c|c||c|c|c||c|c|c|}
\hline
$J_z$ & $\Delta t$(57) & $\Delta t$(115) &
$J_z$ & $\Delta t$(57) & $\Delta t$(115) &
$J_z$ & $\Delta t$(57) & $\Delta t$(115) &
$J_z$ & $\Delta t$(57) & $\Delta t$(115) &
$J_z$ & $\Delta t$(57) & $\Delta t$(115) \\
\hline
1.1 & 0.0591 & 0.0352 & 2.1 & 0.0603 & 0.0319 & 3.1 & 0.0596 & 0.0334 & 4.1 & 0.0678 & 0.0351 & 5.1 & 0.0566 & 0.0350 \\
1.2 & 0.0569 & 0.0354 & 2.2 & 0.0537 & 0.0353 & 3.2 & 0.0540 & 0.0345 & 4.2 & 0.0619 & 0.0324 & 5.2 & 0.0598 & 0.0345 \\
1.3 & 0.0548 & 0.0337 & 2.3 & 0.0559 & 0.0355 & 3.3 & 0.0624 & 0.0363 & 4.3 & 0.0619 & 0.0362 & 5.3 & 0.0611 & 0.0334 \\
1.4 & 0.0582 & 0.0334 & 2.4 & 0.0564 & 0.0345 & 3.4 & 0.0579 & 0.0368 & 4.4 & 0.0575 & 0.0351 & 5.4 & 0.0551 & 0.0351 \\
1.5 & 0.0579 & 0.0335 & 2.5 & 0.0570 & 0.0334 & 3.5 & 0.0579 & 0.0349 & 4.5 & 0.0589 & 0.0332 & 5.5 & 0.0618 & 0.0340 \\
1.6 & 0.0622 & 0.0361 & 2.6 & 0.0550 & 0.0333 & 3.6 & 0.0580 & 0.0363 & 4.6 & 0.0618 & 0.0345 & 5.6 & 0.0613 & 0.0327 \\
1.7 & 0.0593 & 0.0356 & 2.7 & 0.0585 & 0.0330 & 3.7 & 0.0600 & 0.0349 & 4.7 & 0.0591 & 0.0362 & 5.7 & 0.0601 & 0.0342 \\
1.8 & 0.0593 & 0.0351 & 2.8 & 0.0568 & 0.0337 & 3.8 & 0.0591 & 0.0327 & 4.8 & 0.0673 & 0.0349 & 5.8 & 0.0659 & 0.0367 \\
1.9 & 0.0590 & 0.0310 & 2.9 & 0.0584 & 0.0359 & 3.9 & 0.0596 & 0.0348 & 4.9 & 0.0609 & 0.0341 & 5.9 & 0.0636 & 0.0325 \\
2.0 & 0.0551 & 0.0328 & 3.0 & 0.0596 & 0.0335 & 4.0 & 0.0602 & 0.0351 & 5.0 & 0.0578 & 0.0344 & 6.0 & 0.0681 & 0.0358 \\
\hline
\end{tabular}
\caption{SKQD time step used for each $J_z$ and system sizes $n=57$ and $n=115$, rounded to three significant figures.}
\label{supp:tab:dts}
\end{table}

Overall the converged $\Delta t$ are quite uniform: for $n=57$ it ranges between 0.0537--0.0681 and for $n=115$ between 0.0310--0.0368. For comparison, $\frac{\pi}{|H|}$ for $J_z=3.0$ is 0.0393 for $n=57$ and 0.0190 for $n=115$.

\subsection{Configuration recovery}

As described in the main text, we initialize \gls{skqd} to start in the N\'eel state and simulate time evolution from there. Specifically, a two-color bi-partition of the $n=57$ heavy-hex lattice produces an initial state with 33 spin-up particles corresponding to $|0\rangle$ in computational basis and 24 spin-down particles corresponding to $|1\rangle$. For $n=115$ it leads to 67 and 48 respectively. Furthermore, in the \gls{afm} phase, any basis state of the ground state will also need to respect this symmetry.

Although time evolution under our Hamiltonian preserves the number of spin-up and spin-down particles, due to the presence of noise on the quantum computer it is possible to measure bitstrings with a different net spin. From an average of $972759 \pm 10610$ ($999531\pm471$) unique bitstrings sampled for $n=57$ ($n=115$), only an average of $81862\pm8114$ ($81974\pm8138)$ had the correct net spin and were valid bitstrings in the \gls{afm} sector. Adding the invalid bitstrings to the Krylov subspace does not lower the energy and only increases the classical computational cost for subsequent steps. Therefore, we consider two strategies to addressing this. The first is to simply post select and filter out the invalid bitstrings. For the second strategy, on each invalid bitstring we apply bit flips to the spin species which has an excess. For example, if an $n=57$ bitstring has a Hamming weight of 35, we randomly change 2 of the 1 bits to be 0. 

\begin{figure}[ht]
    \centering
    \includegraphics[width=0.85\linewidth]{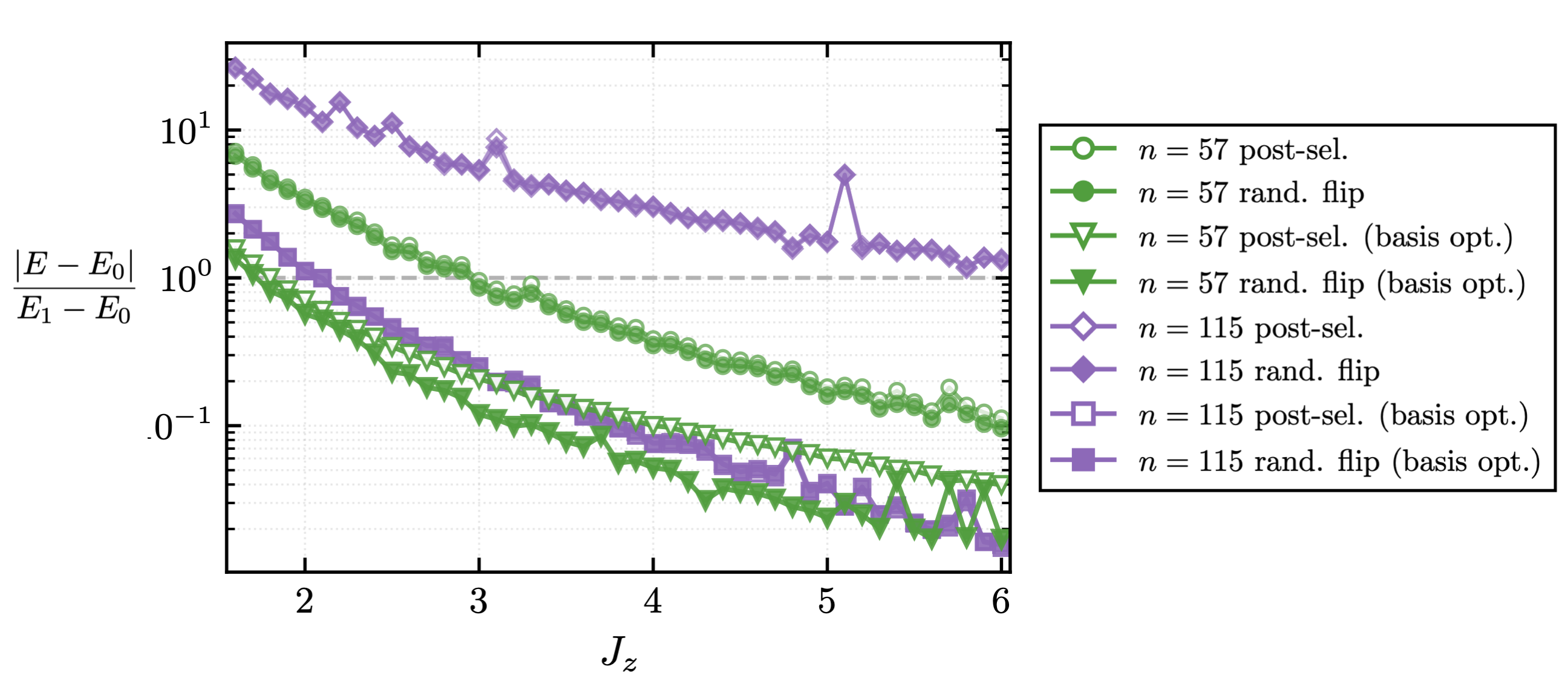}
    \caption{Energy error $(|E-E_0|)/(E_1-E_0)$ relative to the gap to the first excited state as calculated by DMRG. The open markers indicate using post-selection or random bit flipping as configuration recovery. }
    \label{supp:fig:energy_super_comparison}
\end{figure}

Fig.~\ref{supp:fig:energy_super_comparison} shows the difference in energy between the two methods. Here open symbols correspond to post-select whilst filled symbols correspond to random flipping. Looking first at the square and circle markers, we can see that random flipping provides a visible but small decrease in the energy, particularly for large $J_z$. Specifically, this is before basis optimization is applied, which is described in more detail in Supplementary~\ref{supp:subsec:basis_optimization}. Importantly however, looking at the triangle markers, we see that after basis optimization the energy for $n=57$ is notably lower with up to a factor of 2.8 reduction. This implies that the additional bitstrings added by the configuration recovery provides a better subspace in which to optimize the Hamiltonian basis. Finally, for $n=115$ we see no significant improvement in energy, although for consistency we present results using the bit flipped data in the main text.

We note that for other Hamiltonians, physical insights can give a more advanced method of configuration recovery and define a different distribution by which to apply bit flips~\cite{yu2025quantum, piccinelli2025quantum}. Developing a similar scheme for the $XXZ$ model remains an open problem that may yield better data for our scheme.

\subsection{Basis optimization}
\label{supp:subsec:basis_optimization}

After executing the SKQD procedure, we obtain a set of $L$ computational-basis bitstrings spanning a truncated subspace
\begin{equation}
\mathcal{S}=\mathrm{span}(\mathcal{B}),\qquad
\mathcal{B}=\{\,\ket{b_j}\,\}_{j=1}^{L},\qquad
\ket{b_j}\in\{\ket{0},\ket{1}\}^{\otimes n}.
\end{equation}
We project the Hamiltonian $H$ onto $\mathcal{S}$ to form the reduced operator
\begin{equation}
(H_{\mathcal{B}})_{jk}
=\bra{b_j}H\ket{b_k},
\qquad j,k\in\{1,\dots,L\}.
\label{eq:Hproj}
\end{equation}
We then diagonalize $H_{\mathcal{B}}$ to obtain a Rayleigh--Ritz approximation to the ground state within the sampled subspace,
\begin{equation}
\ket{\tilde{\psi}}
=
\sum_{j=1}^{L}\alpha_j \ket{b_j},
\qquad
\boldsymbol{\alpha}\in\mathbb{C}^{L},\;\;
\|\boldsymbol{\alpha}\|_2=1.
\label{eq:psi-tilde}
\end{equation}

Here the projection and diagonalization is done using the built-in functionality of the \texttt{qiskit-addon-sqd} package~\cite{qiskit-addon-sqd}. While $\ket{\tilde{\psi}}$ is optimal within the truncated subspace $\mathcal{S}$, its accuracy as an approximation to the true ground state is limited by the choice of basis $\mathcal{B}$. The SKQD procedure samples only computational-basis bitstrings, so the resulting subspace is inherently biased toward states that are sparse in that representation. If the true ground state has substantial weight on configurations that are not efficiently captured by the sampled strings, then important components of the wavefunction may lie outside $\mathcal{S}$. Consequently, the truncated space may fail to reproduce the relevant low-energy structure even when $L$ is moderately large. To reduce this basis-induced limitation, we introduce a variational basis optimization that finds a representation in which the fixed subspace $\mathcal{S}$ provides a more accurate description of the low-energy physics.

Here we treat the state obtained by SKQD as fixed and adopt a Heisenberg-picture approach, whereby transformations are applied to the operators rather than the state. We then variationally transform the Hamiltonian, equivalently rotating the basis, before diagonalization to obtain a lower-energy approximation. Specifically, we apply a parameterized unitary $U(\boldsymbol{\theta})$,
\begin{equation}
H'(\boldsymbol{\theta})
=
U(\boldsymbol{\theta})\,H\,U^\dagger(\boldsymbol{\theta}),
\label{eq:Hprime-def}
\end{equation}
where $U(\boldsymbol{\theta})$ is a shallow hardware-efficient ansatz. Since $H'$ is related to $H$ by a unitary similarity transformation, the two operators are isospectral. The objective is therefore not to alter the spectrum, but to rotate the operator so that the fixed truncated basis $\mathcal{B}$ gives better estimate of the ground state energy.

We define the basis-optimization cost function as the expectation value of the rotated Hamiltonian in the SKQD state,
\begin{equation}
C(\boldsymbol{\theta})
=
\bra{\tilde{\psi}}H'(\boldsymbol{\theta})\ket{\tilde{\psi}}
=
\bra{\tilde{\psi}}U(\boldsymbol{\theta})\,H\,U^\dagger(\boldsymbol{\theta})\ket{\tilde{\psi}}.
\label{eq:cost}
\end{equation}

Because $\ket{\tilde{\psi}}\in\mathcal{S}$, the cost function can be evaluated entirely from matrix elements of $H'(\boldsymbol{\theta})$ restricted to $\mathcal{B}$. Let $\boldsymbol{\alpha}$ denote the coefficient vector of $\ket{\tilde{\psi}}$. The rotated projected Hamiltonian is defined as,
\begin{equation}
\bigl(H'_{\mathcal{B}}(\boldsymbol{\theta})\bigr)_{jk}
=
\bra{b_j}H'(\boldsymbol{\theta})\ket{b_k}
=
\bra{b_j}U(\boldsymbol{\theta})\,H\,U^\dagger(\boldsymbol{\theta})\ket{b_k}.
\label{eq:Hprime-proj}
\end{equation}
The objective then reduces to the quadratic form,
\begin{equation}
C(\boldsymbol{\theta})
=
\boldsymbol{\alpha}^\dagger
\,H'_{\mathcal{B}}(\boldsymbol{\theta})\,
\boldsymbol{\alpha}.
\label{eq:cost-quadratic}
\end{equation}
In practice, $H'_{\mathcal{B}}(\boldsymbol{\theta})$ is obtained by constructing the rotated operator $H'(\boldsymbol{\theta})$ (e.g., via a Heisenberg propagation or backpropagation procedure) and classically evaluating the projected matrix elements in Eq.~\eqref{eq:Hprime-proj} and Eq.~\eqref{eq:cost-quadratic}.

Let $\boldsymbol{\theta}^\star$ denote an approximate minimizer of Eq.~\eqref{eq:cost}. Although $H$ and $H'(\boldsymbol{\theta}^\star)$ share the same exact ground energy, the vector $\ket{\tilde{\psi}}$ is not guaranteed to be the ground state of the rotated operator when restricted to $\mathcal{S}$. We therefore perform a second Rayleigh--Ritz step by diagonalizing the optimized projected Hamiltonian,
\begin{equation}
H'_{\mathcal{B}}(\boldsymbol{\theta}^\star)\,\boldsymbol{\gamma}
=
E'\,\boldsymbol{\gamma},
\qquad
\ket{\psi'}
=
\sum_{j=1}^{L}\gamma_j\ket{b_j},
\qquad
\|\boldsymbol{\gamma}\|_2=1,
\label{eq:redig}
\end{equation}
where $E'$ is the smallest eigenvalue within the subspace.

By the variational principle restricted to $\mathcal{S}$,
\begin{equation}
E'
=
\min_{\substack{\ket{\phi}\in\mathcal{S}\\ \langle\phi|\phi\rangle=1}}
\bra{\phi}H'(\boldsymbol{\theta}^\star)\ket{\phi}
\le
\bra{\tilde{\psi}}H'(\boldsymbol{\theta}^\star)\ket{\tilde{\psi}}
=
C(\boldsymbol{\theta}^\star).
\label{eq:variational-ineq}
\end{equation}
Hence, re-diagonalization in the optimized basis produces a state $\ket{\psi'}\in\mathcal{S}$ whose energy is no worse than the optimized cost and, in general, strictly improves upon $\ket{\tilde{\psi}}$ whenever the latter is not already variationally optimal for the rotated operator within the truncated space.

\subsubsection{Implementation}

The choice of the unitary transformation $U(\boldsymbol{\theta})$ is heuristic and motivated by empirical performance. We explored several parameterized circuit architectures with varying depths and numbers of parameters. Among these, the circuit that most effectively reduced the energy consisted of a simple two-qubit entangling layer: a CNOT gate between qubits 0 and 1, followed by a single-qubit rotation $R_y(\theta)$ on the target qubit, and a final CNOT gate, i.e.,
\[
\begin{quantikz}
   \lstick{$q_0$} & \ctrl{1} & \qw & \ctrl{1} & \qw \\
   \lstick{$q_1$} & \targ{}  & \gate{R_y(\theta)} & \targ{} & \qw
\end{quantikz}
\]
A single layer of the ansatz is constructed by tiling this two-qubit block across the device according to the optimal three-coloring of the heavy-hex connectivity graph, ensuring that all gates within a layer act on disjoint qubit pairs and can therefore be applied in parallel.
We find that this shallow entangling circuit provides significant improvement in the variational energy within the truncated subspace $\mathcal{S}$. The optimal parameter $\theta$ is determined by minimizing the cost function 
\begin{equation}
C(\boldsymbol{\theta}) = \boldsymbol{\alpha}^\dagger H'_{\mathcal{B}}(\boldsymbol{\theta}) \boldsymbol{\alpha},
\end{equation}
where $H'_{\mathcal{B}}(\boldsymbol{\theta})$ is the rotated projected Hamiltonian defined in Eq.~\eqref{eq:Hprime-proj} and $\boldsymbol{\alpha}$ are the coefficients of the state $\ket{\tilde{\psi}}$ in the sampled basis $\mathcal{B}$ (Eq.~\eqref{eq:psi-tilde}). Here we specifically calculate $H'(\boldsymbol{\theta})$ by building the circuit implementing $U(\boldsymbol{\theta})$ and applying it and its conjugate using \gls{obp}~\cite{fuller2026improved}. Furthermore, we also use \gls{obp} to analytically evaluate the derivative $\partial C/\partial \theta_i$.

For the \gls{skqd} runs employing bit flip configuration recovery, the full bitstring matrix contains approximately 1 million unique configurations, making direct minimization of $C(\boldsymbol{\theta})$ computationally expensive. To address this, we first perform a local optimization over the 100 bitstrings with the largest probabilities, while periodically evaluating the global cost using the full set of bitstrings to monitor convergence. Once the global cost ceases to decrease within a tolerance of $10^{-8}$, we use the parameters obtained from the local optimization as the initial guess for a full optimization over all bitstrings. The local and global optimization steps are repeated for up to 30 iterations and may terminate earlier if the tolerance criterion is not satisfied or if the global cost increases following a local optimization step. Finally, we diagonalize the projected Hamiltonian $H'_{\mathcal{B}}(\boldsymbol{\theta}^\star)$ to obtain the optimized wavefunction $\ket{\psi'}$ in the new basis, which provides the lowest achievable energy within the truncated subspace.

Fig.~\ref{supp:fig:energy_super_comparison} shows the energy error of the \gls{skqd} ground state with respect to the gap. Comparing the filled purple diamonds to the filled purple squares, and the filled green circles to the filled green triangle, shows before and after basis optimization for each system size. Overall the difference to \gls{dmrg} as a ratio to the spectral gap is decreased by an order of magnitude or greater across the phase. Notably, without basis optimization, for $n=115$ none of the ground states would be within the spectral gap, indicating a poor quality solution.

\section{Generated ground states vs DMRG}
\label{supp:sec:skqd_vd_dmrg}

\subsection{Norm distance}

From Lemma 1 of Ref.~\cite{yu2025quantum}, we can compute an upper bound between the \gls{skqd} solution and the ground state in 2-norm distance

\begin{equation}
    \lVert \lvert \psi \rangle - \lvert \phi_0 \rangle \rVert^2
< 2 \left( 1 - \sqrt{ 1 - \frac{E - E_0}{E_1 - E_0} } \right).
\end{equation}

\begin{figure}[h]
    \centering
    \includegraphics[width=0.6\linewidth]{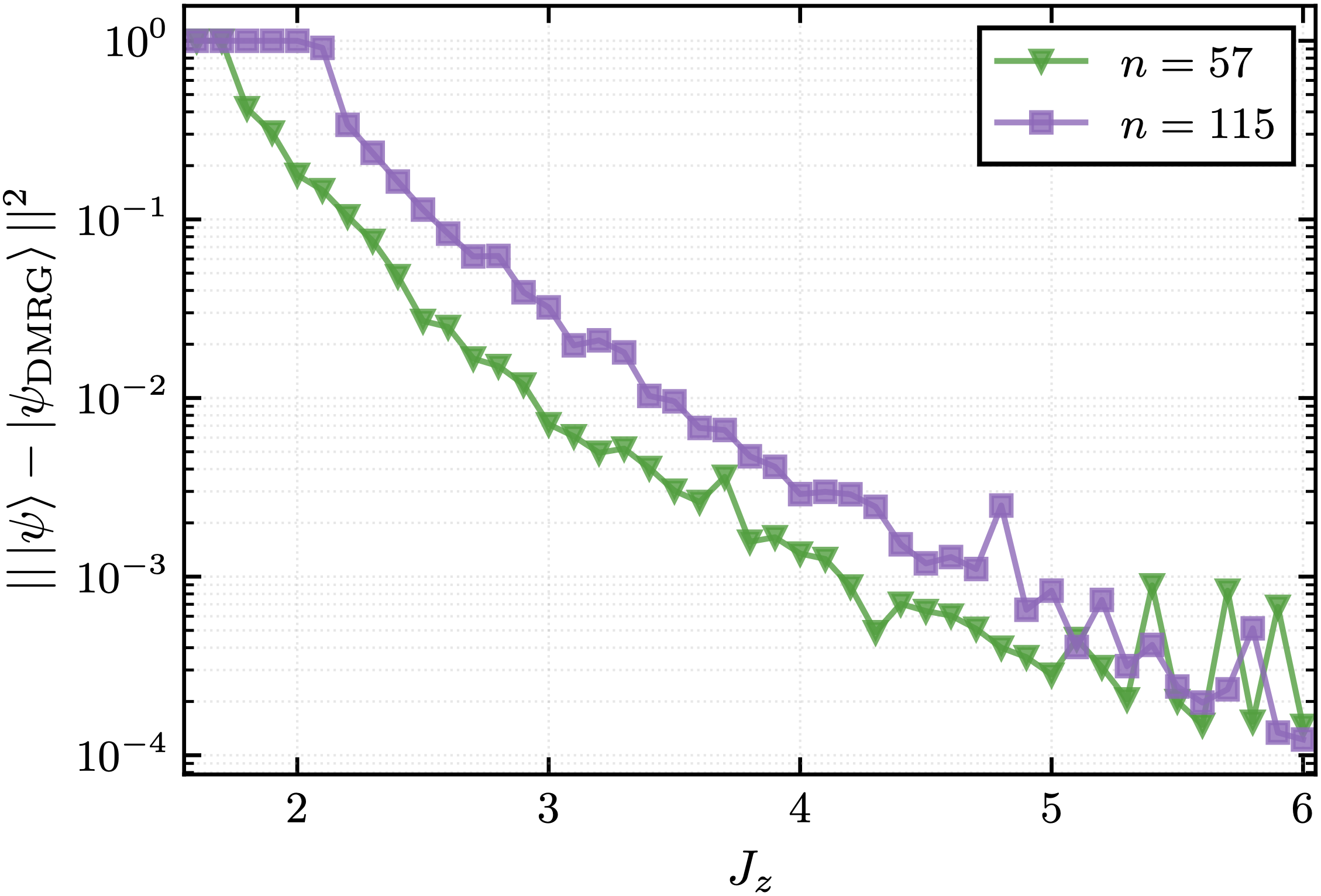}
    \caption{Upper bound on the 2-norm distance between the SKQD and DMRG ground states.}
    \label{supp:fig:norm_distance}
\end{figure}

Fig.~\ref{supp:fig:norm_distance} shows the computed bound for the ground states obtained using random bit flip configuration recovery and basis optimization, which is the data used by the machine learning model. For points where $E$ is larger than the spectral gap, the distance is saturated at 1. However, for larger $J_z$ where the ground state energy is lower we see increasingly small maximum norm distance. This motivates the use of the energy gap as a physically meaningful diagnostic, as when the energy lies below the gap we can make guarantees as to the closeness of the state to the true ground state. While this bound does not directly control local observables, it complements the explicit observable benchmarking presented in the main text and supports that our workflow produces, particularly for large $J_z$, good representations of the true ground states.

\subsection{Observables}

\begin{figure}[h]
    \centering
    \includegraphics[width=\linewidth]{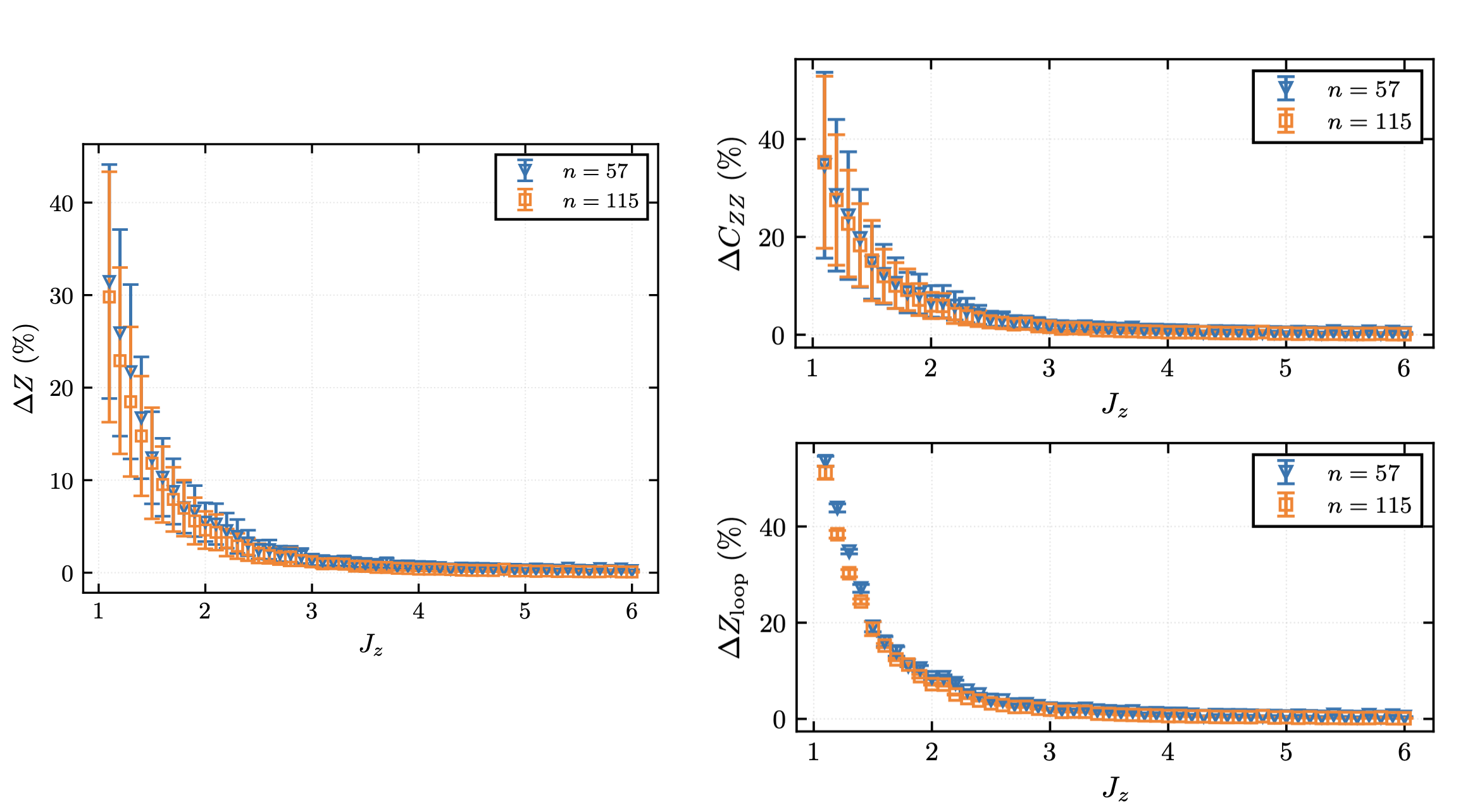}
    \caption{Mean percentage error between quantum data derived from SKQD and DMRG for (left) $Z$, (top right) $C_{zz}$ and (bottom right) $Z_{\mathrm{loop}}$. Here the average is over the individual relative errors across the lattice and error bars show one standard deviation across the distribution of relative errors.}
    \label{supp:fig:skqd_vs_ml_obs}
\end{figure}

Here we expand on Fig.2 of the main text to show the relative error between the generated quantum data and \gls{dmrg} for all 3 observables we later train the model on. Moreover, even though the model is only trained on data $1.6 \leq J_z \leq 6.0$, for completeness we show the relative errors of additional data points in $1.1 \leq J_z \leq 1.5$ generated as the test set for our later experiments extrapolating towards the phase boundary. The results are shown in Fig.~\ref{supp:fig:skqd_vs_ml_obs}. Here the mean relative error is defined such that for the $Z$ observable, 
\begin{equation}
    \Delta Z(\%) = \frac1n\sum_i^n \frac{|\langle Z_i\rangle_{\mathrm{data}} - \langle Z_i\rangle_{\mathrm{DMRG}}|}{\langle Z_i\rangle_{\mathrm{DMRG}}}.
\end{equation}

Similarly $C_{zz}$ is averaged over all 1st and 2nd nearest neighbors, whilst $Z_{\mathrm{loop}}$ is averaged over all 12-qubit loops in the graph.

\section{Machine learning models and training}
\label{supp:sec:deep_learning}

\subsection{Neural Network Architectures}
We describe the neural network architectures used to learn the relationship between ground-state physical observables and the system parameters. We consider two types of architecture: (i) a lightweight per-observable neural network and (ii) a geometry-aware local neural network defined on the heavy-hex lattice. We label these as architecture 1 and architecture 2 respectively. The former does not incorporate information about the underlying lattice geometry and instead relies solely on the global coupling parameter. In contrast, the latter explicitly incorporates the geometric structure of the heavy-hex lattice through the construction of local regions associated with each observable.
\begin{figure*}[ht]
    \centering
    \includegraphics[width=0.65\linewidth]{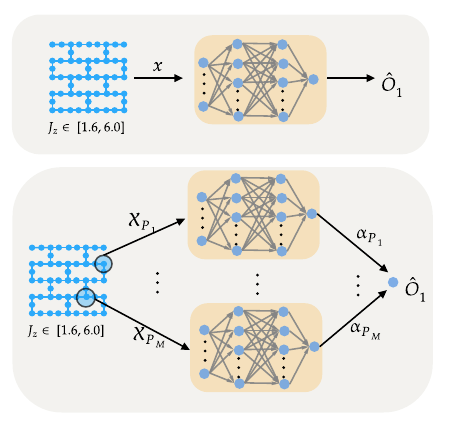}
    \caption{Neural network architectures for observable prediction. (Top) Lightweight per-observable neural network known as architecture 1. The input vector is constructed from a uniform coupling parameter $j_z$ assigned to all edges of the heavy-hex lattice, $\boldsymbol{\chi} = (j_z, \dots, j_z)$, and each observable is predicted independently using a multilayer perceptron. (Bottom) Geometry-aware local neural network known as model 2. The heavy-hex lattice is decomposed into local regions $\{P_i\}$ defined by the lattice geometry and the qubits associated with the observable. For locality parameter $\delta = 0$, each region is restricted to edges incident on these qubits. The corresponding local inputs $\boldsymbol{\chi}_{P_i}$ are processed by dedicated neural networks, and the observable is obtained via a linear aggregation. Each observable is modeled using a separately trained network.}
    \label{fig:ML_model}
\end{figure*}

\subsubsection{Lightweight per-observable neural network}

As illustrated in Fig.~\ref{fig:ML_model}, we consider a minimal architecture in which each observable is modeled as a function of a single scalar coupling parameter $j_z$. The input to the model is therefore a scalar $j_z \in \mathbb{R}$. Each observable $O$ is modeled by an independent multilayer perceptron $f$, yielding
\[
\hat{O} = f(j_z).
\]
The function $f$ is parameterized by a feedforward neural network consisting of fully connected layers and nonlinear activation functions. For a set of observables $\{O_o\}$, independent networks $\{f_o\}$ of identical architecture are instantiated, each producing a scalar prediction. This architecture does not explicitly incorporate spatial information and instead learns a direct mapping from the global coupling parameter to the observable.

\subsubsection{Geometry-aware local neural network}

In the second model, we adopt the architecture introduced in Ref.~\cite{wanner2024predicting} due to its proven scalability and prominence in the literature. In this framework, the prediction of observables is structured around the locality of the Hamiltonian and the decomposition of observables into sums of local operators. Following Ref.~\cite{wanner2024predicting}, we construct a collection of overlapping local regions $\{P_i\}_{i=1}^{M}$, where each region is determined by the heavy-hex lattice geometry and the qubits associated with the observable. Notably, in our application of this model to the spin-\(\frac12\) XXZ Hamiltonian, the anisotropy parameter $J_z$ is uniform across the lattice, so each input edge has the same value. We propose that this encodes a signal of spatial invariance to the model, although further investigation is required into the possible benefit of this.

In this work, we consider a locality parameter $\delta = 0$, such that each region is restricted to the immediate neighborhood of the relevant qubits, without further expansion across the graph. Each region $P_i$ therefore consists of edges incident on the qubits defining the observable, with the corresponding coupling parameters forming the local input $\boldsymbol{\chi}_{P_i}$.

Each local input $\boldsymbol{\chi}_{P_i}$ is processed by a dedicated neural network $f_{P_i}$, implemented as a multilayer perceptron, which produces a scalar feature,
\[
f_{P_i} : \mathbb{R}^{|P_i|} \rightarrow \mathbb{R}.
\]
These local networks act as nonlinear function approximators over restricted regions of the input, capturing local dependencies induced by the geometry of the system. The final prediction is obtained by a linear aggregation of the local outputs,
\[
\hat{O} = \sum_{i=1}^{M} \alpha_{P_i} \, f_{P_i}(\boldsymbol{\chi}_{P_i}),
\]
where $\alpha_{P_i}$ are trainable coefficients. These coefficients can be interpreted as learned approximations to the corresponding coefficients in a Pauli decomposition of the observable. This procedure is repeated independently for each observable. A key advantage of this formulation, as emphasized in Ref.~\cite{wanner2024predicting}, is that it enables scalable learning of observables, as the model operates on local inputs and aggregates their contributions. 

\subsection{Cross validation and training}

We describe the \gls{cv} procedure used for hyperparameter selection and model training, which helps mitigate overfitting to a particular train–validation split, particularly when the dataset is limited. We consider training data in the parameter range $1.6 \leq J_z \leq 6.0$, which is partitioned into three strata: $[1.6, 2.0]$, $[2.1, 4.0]$, and $[4.1, 6.0]$, corresponding to ground states with progressively larger spectral gaps. A stratified train--test split is performed based on the $J_z$ values, with $20\%$ of the data held out for testing. The remaining $80\%$ of the data is used for \gls{cv}.

Hyperparameters are selected using stratified $k$-fold \gls{cv} with $k = 5$, where the distribution of $J_z$ values is preserved across folds. For each hyperparameter configuration, models are trained on four folds and evaluated on the remaining fold, with each fold used once as the validation set. The best validation loss achieved during training is recorded for each fold, and performance is summarized by the mean and standard deviation across folds. Within each fold, we train a neural network model, with the number of outputs determined by the observable under consideration. Models are optimized using the Adam optimizer with mean squared error (MSE) loss. Training is performed for 300 epochs using a batch size of 32, and the best validation loss across epochs is selected after training. All experiments are performed with a fixed random seed for reproducibility.

For architecture 1, we perform a grid search over the network width $\{8, 16, 32, 64, 128, 256\}$, depth $\{2, 3, 4, 5\}$, and learning rate $\{10^{-3}, 10^{-4}\}$, using a \texttt{tanh} activation function and no dropout. Architecture 2 searchers over the same configurations, except lesser network widths $\{8, 16, 32, 64\}$ since it has many more trainable parameters for a given network width. This results in 30 hyperparameter configurations per observable for architecture 1 and 24 for architecture 2. The configuration with the lowest mean validation loss across folds is selected. Tables~\ref{supp:tab:hyperparameters1} and ~\ref{supp:tab:hyperparameters2} show the best configuration found for architectures 1 and 2 respectively.

\begin{table}[htbp]
\centering
\caption{Best hyperparameters found by 5-fold cross validation for architecture 1.}
\label{supp:tab:hyperparameters1}
\begin{tabular}{lcccccc}
\toprule
Dataset & $\langle Z_i \rangle$ ($n=57$) & $\langle Z_i \rangle$ ($n=115$) & $\langle Z_i Z_j \rangle$ ($n=57$) & $\langle Z_i Z_j \rangle$ ($n=115$) & $\langle Z_{\text{loop}} \rangle$ ($n=57$) & $\langle Z_{\text{loop}} \rangle$ ($n=115$) \\
\midrule
Width & 256 & 256 & 256 & 256 & 128 & 256 \\
Depth & 5 & 5 & 5 & 5 & 5 & 5 \\
Learning rate & $1 \times 10^{-3}$ & $1 \times 10^{-3}$ & $1 \times 10^{-3}$ & $1 \times 10^{-3}$ & $1 \times 10^{-3}$ & $1 \times 10^{-3}$ \\
\# Parameters & 15,044,409 & 30,352,755 & 39,854,487 & 83,668,029 & 265,732 & 3,167,244 \\
\midrule
\multicolumn{7}{l}{\textit{Validation Loss}} \\
Mean & $1.24 \times 10^{-5}$ & $7.60 \times 10^{-6}$ & $1.78 \times 10^{-5}$ & $9.37 \times 10^{-6}$ & $1.14 \times 10^{-5}$ & $9.24 \times 10^{-6}$ \\
Std & $5.91 \times 10^{-6}$ & $3.14 \times 10^{-6}$ & $7.84 \times 10^{-6}$ & $4.46 \times 10^{-6}$ & $4.45 \times 10^{-6}$ & $3.49 \times 10^{-6}$ \\
\bottomrule
\end{tabular}
\end{table}

\begin{table}[htbp]
\centering
\caption{Best hyperparameters found by 5-fold cross validation for architecture 2.}
\label{supp:tab:hyperparameters2}
\begin{tabular}{lcccccc}
\toprule
Dataset & $\langle Z_i \rangle$ ($n=57$) & $\langle Z_i \rangle$ ($n=115$) & $\langle Z_i Z_j \rangle$ ($n=57$) & $\langle Z_i Z_j \rangle$ ($n=115$) & $\langle Z_{\text{loop}} \rangle$ ($n=57$) & $\langle Z_{\text{loop}} \rangle$ ($n=115$) \\
\midrule
Width & 64 & 32 & 32 & 32 & 32 & 32 \\
Depth & 5 & 5 & 4 & 4 & 3 & 3 \\
Learning rate & $1 \times 10^{-3}$ & $1 \times 10^{-3}$ & $1 \times 10^{-3}$ & $1 \times 10^{-3}$ & $1 \times 10^{-3}$ & $1 \times 10^{-3}$ \\
\# Parameters & 124,214,400 & 134,100,120 & 64,961,408 & 281,275,368 & 1,180,160 & 7,302,240 \\
\midrule
\multicolumn{7}{l}{\textit{Validation Loss}} \\
Mean & $1.26 \times 10^{-5}$ & $6.49 \times 10^{-6}$ & $1.95 \times 10^{-5}$ & $9.34 \times 10^{-6}$ & $1.40 \times 10^{-5}$ & $1.04 \times 10^{-5}$ \\
Std & $6.48 \times 10^{-6}$ & $2.82 \times 10^{-6}$ & $8.75 \times 10^{-6}$ & $4.31 \times 10^{-6}$ & $5.94 \times 10^{-6}$ & $4.08 \times 10^{-6}$ \\
\bottomrule
\end{tabular}
\end{table}

Using the best found hyperparameters from \gls{cv}, we then train the models for 200 epochs. The same 20\% of the data is removed for held-out testing as before, of which the remaining 75\% is used for training and 5\% for validation. The validation data is used define an early-stopping criteria, such that if the validation less has not decreased by $10^{-8}$ in the last 50 epochs, training is halted.

\subsection{Hardware}

Cross validation, model training, and inference were performed on a MacBook Pro with an Apple M2 Max processor (38‑core GPU) and 64GB of unified memory.

\section{Architecture comparison}
\label{supp:sec:ml_vs_skqd}

\subsection{Generalization within the interior of the phase}

\begin{figure}[ht]
    \centering
    \includegraphics[width=\linewidth]{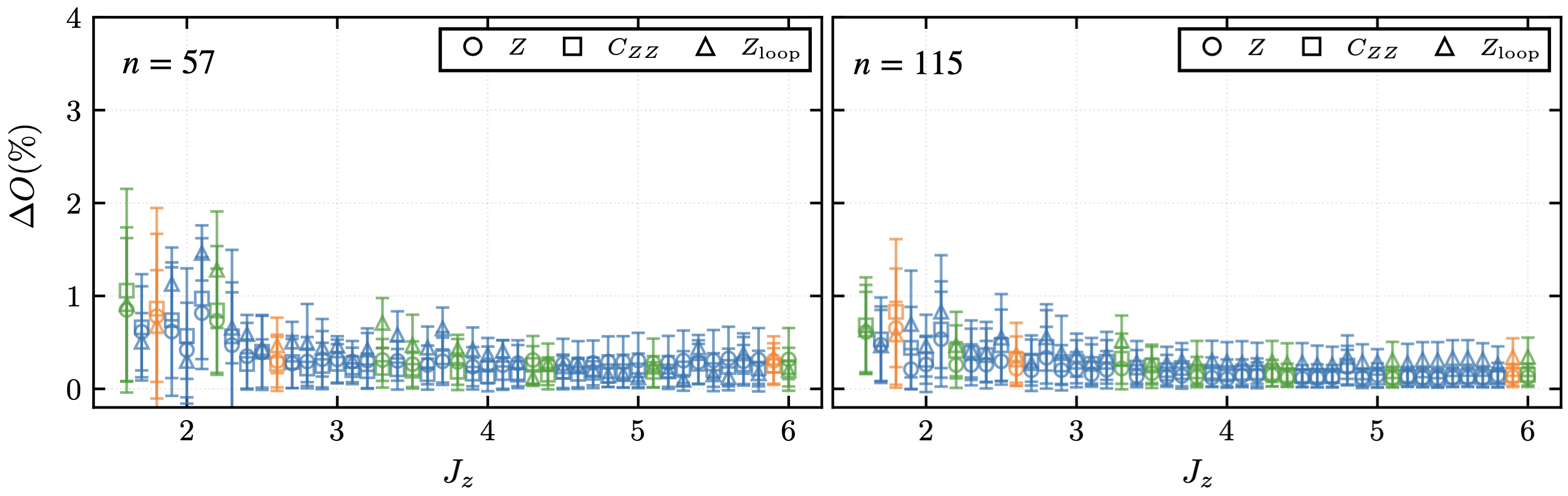}
    \caption{Model generalization using architecture 1 within the interior of the AFM phase as shown by mean relative error $\Delta O_{\textrm{ML, data}}(\%)$. Error bars show one standard deviation of the distribution of individual relative errors. Blue, orange and green markers correspond to $J_z$ points from the train, validation and test data set respectively.}
    \label{supp:fig:model1_interpolation}
\end{figure}

Fig.~\ref{supp:fig:model1_interpolation} shows the average relative error when predicting within the training distribution for architecture 1. Error bars show one standard deviation of the spread of relative error across the observables and the blue, green and orange correspond to $J_z$ points in the train, validation and validation sets respectively. Compared to Fig.~3 of the main text, we see the in-distribution generalization performance of architecture 1 is very similar.

\subsection{Generalization towards phase boundary}

\begin{figure*}[ht]
    \centering
    \includegraphics[width=\linewidth]{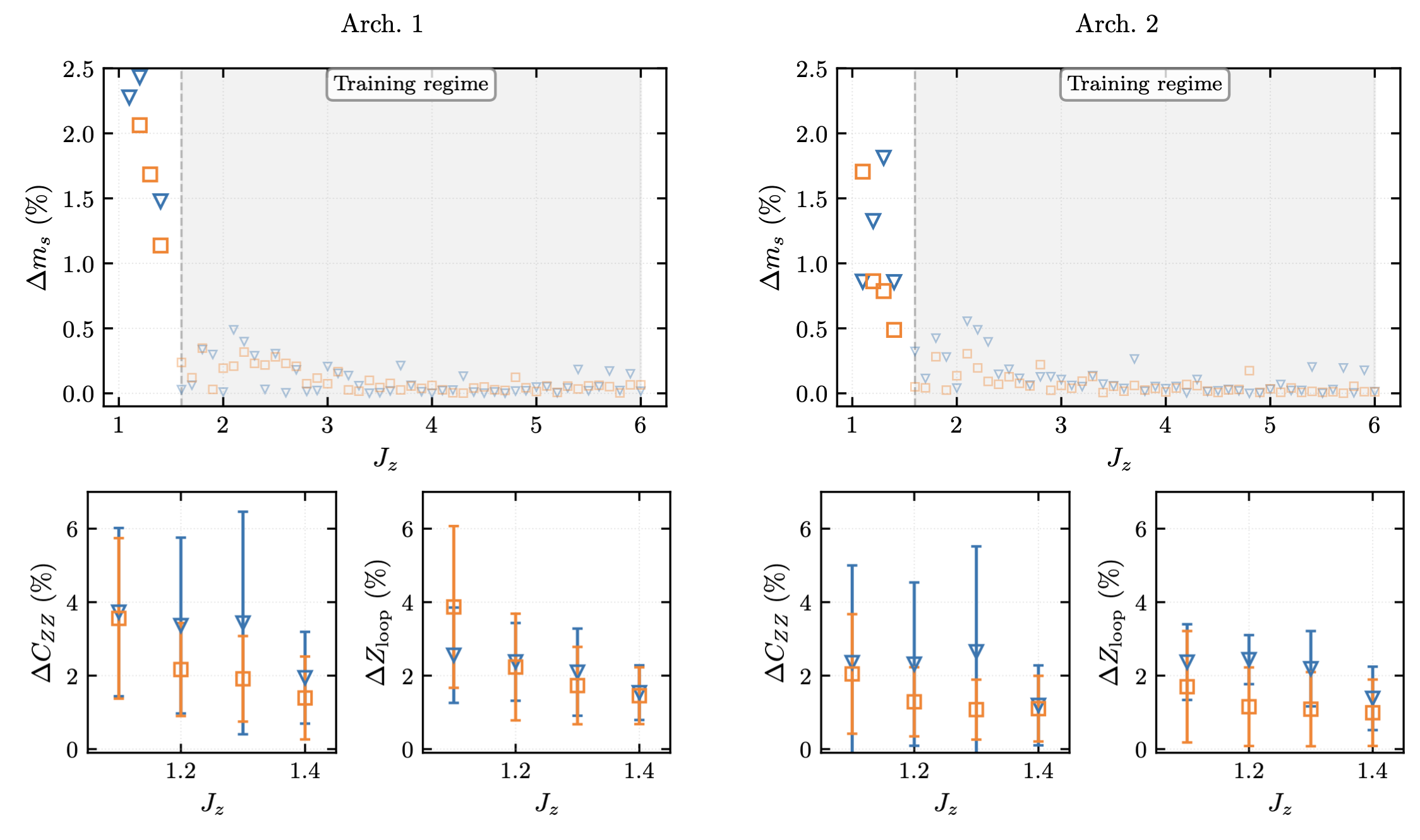}
    \caption{Comparison between architecture 1 and 2 generalization towards the phase boundary. Relative error between local observables predicted by \gls{ml} and data from new quantum experiments outside the training regime.}
    \label{supp:fig:model_comparison_phase_boundary}
\end{figure*}

Fig.~\ref{supp:fig:model_comparison_phase_boundary} shows a comparison in the average relative error in predicting local observables outside of the training distribution and towards the phase boundary. Here the architecture 2 results is the same data as in Fig. 4 of the main text but adjusted to be on the same scale as architecture 1. Here we see clear evidence that architecture 2 predicts the staggered magnetization $m_s$, as well as the two-point correlators and loop correlators, with on average less error. We note that models using architecture 2 contain more parameters that architecture 1, and thus a systematic study between the two whilst controlling for number of parameters is an interesting future direction of research. However, we consider architecture 2 in the main text due to its provable efficiency~\cite{wanner2024predicting} and improved out-of-distribution generalization in these limited benchmarks.

\pagebreak

\bibliography{bibliography}

\end{document}